# Assessing the Utility of Large Language Models for Phenotype-Driven Gene Prioritization in Rare Genetic Disorder Diagnosis

## Author list

Junyoung Kim[1], Jingye Yang[2,4], Kai Wang[2,3], Chunhua Weng[1,*], Cong Liu[1,*]

## Detailed Affiliations

[1]Department of Biomedical Informatics, Columbia University, New York, NY, USA;

[2]Raymond G. Perelman Center for Cellular and Molecular Therapeutics, Children's Hospital of Philadelphia, Philadelphia, PA 19104, USA

[3]Department of Pathology and Laboratory Medicine, University of Pennsylvania, Philadelphia, PA 19104, USA

[4]Department of Mathematics, University of Pennsylvania, Philadelphia, PA 19104, USA

(*: equal-contribution senior corresponding authors)

## Abstract

Phenotype-driven gene prioritization is a critical process in the diagnosis of rare genetic disorders, involving identifying and ranking potential disease-causing genes based on observed physical traits or phenotypes. While traditional approaches heavily rely on curated knowledge graph with phenotype-gene relations, recent advancements in large language models (LLMs) have opened doors to the potential of AI predictions through extensive training on diverse corpora and complex models. In this study, we conducted a comprehensive evaluation of five large language models, including two Generative Pre-trained Transformers (GPT) series, and three Llama2 series, assessing their performance across three key metrics: task completeness, gene prediction accuracy, and adherence to required output structures. We conducted various experiments, exploring various combinations of models, prompts, phenotypic input types, and task difficulty levels. Our findings reveal that even the best performed LLM, GPT-4, only achieved an accuracy of 16.0%, which still lags behind traditional bioinformatics tools. Among the five LLMs, prediction accuracy increased as the parameter/model size increased. A similar increasing trend was observed for the task completion rate (70.7% to 94.2%), with complicated prompts more likely to increase task completeness in models smaller than GPT-4. On the other hand, complicated prompts are more likely to decrease the structure compliance rate, but no prompt effects were observed in GPT-4, which achieved an almost perfect structure compliance rate in all experiments. Comparing to HPO term-based input, LLM was also able to achieve better-than-random prediction accuracy by taking free-text input directly, though the performance is slightly lower than with the HPO input. Bias analysis showed that certain genes, such as MECP4, CDKL5, and SCN1A, are more likely to be top ranked by LLMs, potentially explaining the dramatic variances observed across different datasets. In conclusion, this study provides valuable insights into the integration of LLMs within genomic analysis, contributing to the ongoing discussion on the utilization of advanced LLMs in clinical workflows.

## Introduction

Phenotype-driven gene prioritization is a process that involves identifying and ranking candidate disease-causing genes by examining an individual's observed physical traits, known as phenotypes, in contrast to genotypes. It plays a crucial role in rare disease diagnosis when analyzing genomic data from high-throughput (e.g., Whole Genome/Exome Sequencing) experiments or design virtual panels for diagnosis purposes. It guides the additional experimental or diagnostic efforts toward the most promising candidates[1]. The underlying principle is based on the assumption that disease-related phenotypes are a result of one or more gene dysfunction. By leveraging the established genotype-phenotype-disease associations, phenotype-based analytical models have been developed, resulting in numerous bioinformatics tools. For instance, Phenomizer[2] compares individual phenotypes to a database of known genetic diseases and their associated phenotypes, and it then ranks the potential genetic diseases based on the semantic similarity of the patient's phenotype to known conditions within the Human Phenotype Ontology. Exomiser[3] integrates various data sources, including gene-disease associations, variant databases, and gene ontology data, and employs a random-walk analysis to score and rank candidate genes or variants. Similarly, AMELIE[4] constructed a comprehensive disease-phenotype knowledgebase by integrating various databases and resources, including ClinVar, HGMD, and parsing relationship from literature. It then developed a machine learning classifier to rank candidate genes. Phenolyzer[5] and its successor Phen2Gene[6] integrated Human Phenotype Ontology (HPO) annotations, gene-disease databases, and gene-gene network and apply a probabilistic framework to build a phenotype-driven gene prioritization tool. With the recent advancement in deep learning, network inference methods based on modern deep learning frameworks have also been explored. For example, DeepSVP[7] constructs a graph from ontology axioms and employs DL2Vec approach for gene-phenotype association prediction. CADA[8] utilizes graph embedding techniques to predict links between phenotype and diseases.

The majority of the existing tools rely on established knowledge databases or knowledge graphs that connect phenotypes with genes (or monogenic diseases). These databases are typically curated through either manual processes or mining the literature, which can be prone to errors or lack comprehensiveness. Additionally, most of these bioinformatics tools can only process "term-based" input, requiring natural language processing techniques to parse the terms from clinical notes. We hypothesized that recent advancements in large language models (LLMs), trained on massive and diverse dataset, could potentially provide an end-to-end, text-to-gene solution to this task by leveraging their extraordinary ability to understand natural language and access extensive and diverse repositories of information[9]. These LLMs represent noteworthy advancements in the journey towards AGI, aiming to develop intelligent systems capable of understanding and executing a diverse array of tasks, akin to human-level cognitive capabilities[10]. These models begin by employing a transformer-based decoder architecture to pre-train a base language completion model without supervision. Subsequently, the base model undergoes a fine-tuning process with human feedback and additional refinement through reinforcement learning, guided by a reward model trained using supervised methods, which can lead to the development of a ChatBot. Numerous articles have indicated the potential of LLMs in the medical field, covering a range of applications, including individual education[11], appointment scheduling[12], optimization of clinical decision support systems[13], aid in data collection for clinical studies[14], enhancement of information retrieval in electronic health records (EHRs)[15], and summarization of evidence in publications[16]. Previous studies have shown that the parameter/model size is usually correlated with performance. For example, GPT-4 outperformed GPT-3.5 in biomedical classification tasks, reasoning tasks[17] and accuracy in question and answer tasks[18]. Similarly, Llama 2 models also demonstrated that larger model size correlated with higher performance in biomedical question and answer tasks [19].

In this study, we will particularly focus on two advanced ChatBot-based LLMs: Generative Pre-trained Transformer (GPT) series, including GPT-3.5 (also known as ChatGPT) and GPT-4[20], and the Llama2 series, including Llama2-7b-chat (with 7b indicating 7 billion parameters), Llama2-13b-chat and Llama2-70b-chat[21,22]. We will explore LLMs' ability to analyze human phenotypic presentations and predict genetic diagnoses. Our primary objectives include evaluating LLM's performance in completing the designated task, achieving accurate gene prioritization, and adhering to output structure requirements. We will also assess the impact of various factors, including prompts, model sizes, task difficulty levels, and phenotypic input type. Overall, this study aims to provide a comprehensive assessment of how LLMs can potentially be integrated into the current clinical genomic analysis workflow for rare disease diagnosis.

## Material and methods

### Datasets

We utilized a list of publicly accessible datasets consisting of total 276 de-identified individuals who had been diagnosed with Mendelian diseases[6]. It is worth highlighting that the selection of phenotypes for these individuals was based on the criterion that the diseases were monogenic, and the causal genes were previously established. Data were collected from five distinctive sources as outlined in **Table 1**. There are a total of 165 distinct genes in the final pool of diagnosed genes. This approach allowed us to include a broad spectrum of Mendelian diseases and genes, ensuring that the dataset covered various genetic conditions. An IRB exemption was obtained from the Columbia University Review Board.

For individuals in each dataset, their HPO concepts and final diagnosed genes were previously curated[6]. Free-text phenotypic descriptions were further collected from 53 original articles for 125 individuals. This was accomplished by identifying the relevant section in the articles and manually extracting the original text that provided detailed information about the individual's phenotype presentations. One such example is "At 6 years IQ was evaluated (Verbal IQ 73, Performance IQ 58). On physical examination at 15 years, she has a normal head circumference (+0 SD), facial dysmorphism and thoracolumbar kyphoscoliosis." To avoid data leakage, we excluded genetic/genomic analysis-related text from the original article. Considering earliest developed model included in this study is GPT-3, with a knowledge cutoff in 2021 (i.e., trained before September 2021), we only included articles published before 2021.

### Prompt Engineering

To maximize the utilization of LLMs, effective prompt engineering is essential[23]. A prompt refers to a set of instructions or input provided to a LLM in order to steer its response or generate desired output. Prompt engineering involves the process of designing and formulating prompts to elicit desired responses from language models[24]. Considering that the fundamental function of an LLM is to accomplish language completion tasks, it's rational to assume that the quality, clarity, and specificity of the prompt could significantly impact LLM's performance[25]. To assess how variations in prompts could influence the performance, we formulated various prompts based on the intricate interplay between roles, instructions, and model performance (as shown in **Table 2**): (1) The "Original" prompt serves as the fundamental query to use individual phenotype features for gene prioritization. (2) The "Original + Role" prompt introduces a role assignment component, casting the model in the role of a "genetic counselor". While preserving the core objective of gene prioritization, this prompt integrates the role-based perspective, which is also endorsed as a 'system' message by OpenAI[24]. (3) The "Original + Instruction" prompt included additional guidance sentences for the model, instructing it to utilize its accumulated knowledge without requiring real-time database access for predictive purposes. (4) The "Original + Role + Instruction" prompt integrates the role of a genetic counselor with the previously mentioned instruction, offering the most comprehensive approach. Each input prompt was presented as a zero-shot[26] scenario without providing examples. Besides different prompt engineering techniques, we also considered different factors in constructing the final prompt. We assessed two input categories for individual phenotypic features: those presented through HPO terms and those via narrative descriptions (obtained as described above). In addition, we designed an easy task (i.e., making a correct gene prediction within the top 50 predictions) and a challenging (i.e., within the top 10) task to evaluate the performance.

## Experiment Design

For GPT-series, we utilized the OpenAI API, making API requests with various prompts to retrieve the gene prioritization predictions for each prompt (as shown in **Figure 1**). GPT-3.5-turbo was used for GPT-3 evaluation. We set the temperature parameter to zero to make the predictions more deterministic. In total, there were 32 experiments for each input case by combining various factors including GPT versions (GPT-4 and GPT-3.5), prompts (Original, Original + Role, Original + Instruction, and Original + Role + Instruction), phenotypic features input type (HPO concepts and free-text), and predictive task difficulties (Top 10 and Top 50).

For the Llama2-series, we utilized the Replicate's API (a cloud-based service that provides access to a wide range of pre-built models) to assess the three models with different parameter sizes. To save study costs, we did not consider the prompt effect by only including the most complicated prompts (Original + Role + Instruction) in the experiments. In total, there were 8 experiments for each input case. All experiments were repeated three times (i.e. three LLM response generated independently) to measure the variability of LLMs. A previous report[27] demonstrated the possible time-dependent nature of GPTs. To reduce the bias associated with calendar days and because of the closed-source nature of GPT, we permutated the experiment sequence. The final API execution date was in Aug 2023. For Llama2, we did not have that concern due to its open-source nature.

## Evaluation Metrics

We assessed the performance of the LLMs on three outcomes, including the task completeness of the output, accuracy of generated gene lists, and adherence to specified output structural requirements. The evaluations metrics were specified as follows.

**Task Completeness** measures whether LLMs can produce a gene list for gene prioritization tasks. If fewer than half of the required number of genes were generated (e.g., < 25 out of the top 50 predictions), we considered LLMs did not complete the task. It's important to note that we excluded fabricated genes and counted duplicated genes in the prediction list only once. If LLMs declined to produce prediction results, we also considered it as an incomplete task. Supplementary **Figure 1** presents two instances of GPT responses indicating task incompletions.

**Accuracy of Gene Predictions** evaluates whether the true diagnosed gene is identified within the top (10 or 50) predicted gene list generated by the GPTs. Two sub metrics were designed. First, only experiments that successfully completed the task as mentioned above were considered for accuracy assessment. Second is the overall accuracy, which measures the same as the other two metrics that includes all experiments. For experiments where the task was not completed, we considered the gene prediction results to be incorrect. Supplementary **Figure 2** presents examples of correct or incorrect predictions made by GPT.

**Output Structure Compliance** evaluates whether GPT's response adheres to the specified output format requirement. Given GPT's tendency of providing free-text responses, we deemed the outcome compliant if a portion of the response matched a set of predefined regex patterns. As we instructed GPT to generate "{Not Applicable}" for tasks that couldn't be completed, we could independently evaluate this metric regardless of GPT's task completeness. Supplementary **Figure 3** presents examples of GPT's output being compliant (or non-compliant) with the output structure.

For each experiment $e$, we calculated the task completeness rate (or gene prediction accuracy rate, output structure compliance rate) $p_e^{(t)} = \sum_{m=1}^{M} \frac{I^{(t)}(m,e)}{M}$ for each repetition $t$ $(t = 1,2,3)$, where $M$ is the total number of cases for evaluation in this experiment, and $I^{(t)}(m, e) = 1$ if the task was completed (or gene prediction is accurate, output structure is compliant). The standard deviation for each metric can also be calculated by measuring the variance among three repetitions. It's important to highlight that output format and task completeness are independent of each other, whereas accuracy measurement is exclusively applicable to those experiments where tasks have been successfully completed.

Due to the substantial number of experiments and the significant human effort required to assess the outcomes, we developed an in-house program to automate the measurement of the three aforementioned metrics. To assess whether task was completed, we parsed the responses and compared them against HGNC gene symbols (including previous symbols and alias symbols) using regular expression, while excluding common gene name errors (e.g., SEPT1, MAR1) as detailed in the Gene Name Errors Screen project[28]. This program was also used to measure the accuracy of gene predictions by comparing the responses with the diagnosed gene HGNC symbols (including previous symbols and alias symbols). To evaluate adherence to the output format, we established regular expression patterns based on the prompt's requirements and searched for these patterns within LLMs' responses. The automated program was inspected manually on 249 randomly selected experiments, and a 100% accuracy was achieved.

# Result

## Accuracy of Gene Predictions Among Different Models

**Figure 2A** showed the prediction accuracy rate of LLMs underutilizing "Original + Roles + Instruction" prompt. On average, GPT-4 achieved the highest overall prediction accuracy rate of 15.5%, followed by GPT-3.5 (12.7%), Llama2-70b-chat (6.4%), Llama2-13b-chat (5.6%) and Llama2-7b-chat (3.6%). The accuracy rate among completed tasks showed a similar trend, but with smaller gaps among different models.

**Figure 2B** further demonstrated the performance among different datasets for top 50 tasks. As the best performed LLM, GPT4 displayed the highest average overall accuracy rate of 29.51% (32.24% among completed) in CSH, closely followed by CUMC, (27.16%, 30.03% among completed). In the context of the top 10 tasks, CUMC outperformed the others with the highest accuracy, with CSH in close pursuit. The performance in DGD was noted at 24.71% (27.21% among completed), while the performance in AJHG was only 0.75% (0.89% among completed), and TAF1 exhibited the lowest rate of 0% (since TAF1 gene was never predicted by GPTs). This result deviates from the trends observed in the results from other software applications, where TAF1 typically occupied the top position with close to perfect accuracy, followed by CUMC, DGD, AJHG and CSH. Moreover, this variability across the datasets assumes potential biases in the models' performance.

In general, even the best-performing LLM model still lags behind traditional bioinformatics tools. An example of comparing performance between LLMs and traditional bioinformatics tools was shown in **Figure 2C**, where the accuracy of predicting the top 50 candidate genes in the DGD datasets was plotted.

## Gene-Dependent Prediction Bias Analysis

A further investigation of the GPT-4's response revealed potential gene-dependent bias associated with the LLM models, which might explain the variations in accuracy observed across datasets. **Figure 3** shows the top 10 genes most frequently predicted by GPT-4, with six of them did not appear in the diagnosed gene pool. For example, FOXG1 consistently appeared in GPT's output, with 2125 instances (22.1%) in the top 10 predictions and 2791 instances (29%) within the top 50 predictions, despite never occurred in the diagnosed pool. In contrast, genes that frequently appear in the diagnosed gene pool, such as DHX30, are notably absent from both the top 10 and top 50 predictions among all experiments.

**Table S1** shows the odds ratio calculated as the observed prediction times versus expected prediction times in the dataset for each gene across different models. Overall, 6383 genes never occurred in the final diagnosed pool but predicted by GPTs. Most of them (4693) were generated when the task involved top 50 predictions. Furthermore, considering the 165 genes in the final diagnosed pool given the 'Original + Role + Instruction' prompt, GPT-4 did not predict 104 genes (OR = 0), and GPT-3.5 failed to predict 116 genes. Llama2-70b-chat missed predictions for 135 genes, Llama2-13b-chat for 140 genes, and Llama2-7b-chat for 151 genes. This suggests that GPT models were more likely to make successful predictions. Additionally, the number of genes correctly predicted more than fifty percent of the time (OR > 5) were 38 by GPT-4, 34 by GPT-3.5, 8 by Llama2-7b-chat, 5 by Llama2-13b-chat, and 9 by Llama2-70b-chat.

## Effect of Different Factors in Predicting Accuracy

**Table 3** and **Table 4** showed the gene prediction accuracy rate under different settings in GPTs and Llama2 models. As expected, the gene prediction accuracy for the more challenging task (i.e., making correct predictions within the top 10) was significantly lower (12.11%; 10.15% overall) than the easier task (17.41%; 13.06% overall). However, the differences between these two tasks do not seem as large as expected, which could indicate that LLMs may either make correct predictions in the early sequence of output or cannot make correct predictions at all.

In top 50 tasks (**Table 4**), HPO concept inputs achieved an average accuracy rate of 20.67% for GPT-4 (18.18% overall), whereas free-text inputs had an average accuracy rate of 12.25% (11.67% overall), and GPT-3.5 and Llama2-chat models showed the same tendency. Similar trend observed in top 10 tasks but less discrepancy (**Table 3**). This suggests that despite LLM's ability to understand narratives, structured input can contribute to improving its prediction accuracy.

We did not observe prompts have a significant effect on the accuracy of gene predictions in GPT-4. However, we found that the prompt matters in GPT-3.5, especially for overall accuracy, where there is a tendency for more detailed prompt content to yield better accuracy results. The "Original" prompt yielded 2.41% overall accuracy on average in predicting top 50 genes, with "Original + Role + Instruction" prompt showed 15.3% accuracy, which can be explained by the significant impact of prompts on task completion rate observed in GPT-3.5 (described in the section below). A similar trend was observed in the predictions of the top 10 genes, albeit with less discrepancy.

A breakdown of factor impacts under different combinations of other factors can be found in **Table S2**.

## Task Completeness Among different models

The overall task completion rate was 78.34%, with the average for GPT models at 79.42% and Llama2-chat models (using the 'Original + Role + Instruction' prompt) at 75.45%. Similar to accuracy, larger model sizes tended to achieve higher task completion rates. Specifically, GPT-4 achieved an almost perfect average completion rate of 94.22% across all scenarios, which decreased to 64.62% for GPT-3.5. This trend was also observed in the Llama2-chat models (using the 'Original + Role + Instruction' prompt only), where the 7b-chat model had a completion rate of 70.70%, the 13b-chat model 72.82%, and the 70b-chat model 82.83%.

While prompts had no significant impact on task completion in GPT-4, they significantly influenced GPT-3.5, where more complex prompts tended to achieve better completion rates. As shown in **Table 5**, the completion rate for GPT-3.5 was 23.15% with the simplest prompts, increasing to 93.93% with the most complex prompts.

All LLMs were more likely to complete easier tasks, except Llama2-70b-chat model. When requesting only the top 10 genes, an average completion rate of 83.82% (GPT-4: 98.86%, GPT-3.5: 68.79%, Llama2-7b-chat: 87.11%, Llama2-13b-chat: 91.85%, Llama2-70b-chat: 82.46%) was achieved. However, extending the request to include the top 50 responses resulted in a significantly lower rate of 75.02% (GPT-4: 89.59%, GPT-3.5: 60.45%, Llama2-7b-chat: 54.28%, Llama2-13b-chat: 53.78%, Llama2-70b-chat: 83.21%). Further investigation revealed that the largest discrepancies occurred when GPT-3.5 was tasked with generating the top 50 results, achieving 12.80% task completion on average (**Table S2**).

Similarly, input type also had an impact, though differently across models. GPT models generally had higher completion rates on free-text input; for GPT-4, it ranged from 93.40% (HPO) to 96.03% (free-text), and for GPT-3.5, from 62.79% (HPO) to 68.67% (free-text). However, 2 out of 3 Llama2-chat models had higher completion rates with HPO input: Llama2-7b-chat achieved 71.56% with HPO and 68.60% with free-text, while Llama2-70b-chat had 85.27% with HPO and 77.47% with free-text. Conversely, Llama2-13b-chat showed the opposite trend, with 70.89% for HPO and 77.07% for free-text. Nonetheless, these factors are likely influenced by other variables.

A breakdown of factor impacts under different combinations of other factors can be found in **Table S2**.

## Structure Compliance

**Table 6** demonstrates the output structure compliance rate for various experiments. Across nearly all settings, GPT-3.5 struggled to generate compliant output responses, achieving an average compliance rate of 27.32%. Notably, the average compliance rate significantly drops when employing more complex prompts, approaching zero for all categories that use the "Original + Role + Instruction" prompt structure. In contrast, GPT-4 exhibited a significantly higher and roust output structure compliance rate of 79.28% ($p < 0.01$). Llama2-chat models showed a divergence from GPT models, with the 70B model achieving the lowest structural compliance rate (0.25%) and the 13B model the highest (1%). Input type also impacted compliance, with HPO inputs achieving an average compliance rate of 80.01% with GPT-4, compared to 77.67% for free-text inputs. This discrepancy was more emphasized in GPT-3.5, where HPO concept inputs achieved a 30.15% compliance rate, while free-text inputs achieved only 21.07%. Interestingly, in contrast to the task completeness assessment, the "Original" prompt achieved the highest average compliance rate of 62.97% in GPT-3.5, while additional prompts significantly decreased output structure compliance to 30.59% (with "Role"), 15.46% (with "Instruction"), and 0.25% (with "Role and Instruction"). Notably for GPT-4, the "Original" prompt resulted in the lowest structural compliance rate of 77.64%. Furthermore, the task difficulty level was crucial; in GPT-4, easy tasks (Top 10) achieved 99.98%, while difficult tasks (Top 50) achieved 58.58%. This trend was similarly observed in GPT-3.5, albeit with a smaller discrepancy (31.17% for Top 10 and 23.46% for Top 50).

## Variability analysis of GPT

We conducted a detailed investigation into the variabilities in GPT-generated responses. Among the 6,817 unique experiments analyzed, 401 (of which 4,887 completed the task in at least one iteration) yielded different results (in terms of task completion) across three iterations. In terms of accuracy, out of the 5,135 unique experiments analyzed, 153 (with 697 making accurate predictions in at least one iteration) showed differences in achieving the accurate results across the three iterations. Regarding structure compliance, among the 5,296 settings analyzed, 1,120 (with 2,839 yielding compliant results) exhibited differences in compliance across the three iterations.

# Discussion

Previous studies have consistently shown that LLMs achieved remarkable performance across various medical applications, including the Ophthalmology Exam[31], USMLE Sample Exam[32], and progress notes summarization[33]. A few studies have explored the application of LLM models in the clinical genetic/genomic fields. One study suggested ChatGPT achieved a 68.2% accuracy regarding 85 human genetics-related multiple-choice questions and outperformed human response[34]. Another study found GPT-4 were capable of generating cases with similar names regarding the discovery of gene set function[35]. Transformer-based model has also been applied to gene sequence classification tasks and was found to perform better with fewer training epochs than other machine learning approaches[36]. To the best of our knowledge, our study represents the first comprehensive evaluation of most advanced LLMs' performance for phenotype-based rare genetic disease diagnosis.

Base large language models were initially conceived as end-to-end, next-token prediction models. For this task, ChatBot-like models are required to achieve the prediction without extra fine-tuning. GPT-3.5 trained similarly to InstructGPT and optimized for dialogue from GPT-3[37]. Llama2-chat is also fine-tuned for dialogue usage based on Llama2 starting from Supervised Fine-Tuning(SFT) and trained utilizing Reinforcement Learning with Human Feedback (RLHF)[21]. GPT-4 also utilized RLHF and supports a multimodal model[38]. In fact, when we used the base Llama2 model, it appeared to lack comprehension, repetitively generating the text by repeating the questions and examples mentioned in the prompts (specific prompts were designed with "answer is: " pending as the last few tokens to prompt base models for next token prediction task). This tendency remains consistent, irrespective of the number of genes to be generated or variations in phenotypic descriptions; it always begins with a gene provided in the prompt example (in this case, ABCA1 or ABC1), until a specified token length is satisfied. The accuracy of gene prediction by Llama2 appears to be entirely random. However, once the model is trained in a supervised fashion (GPT-3.5/GPT-4 or Llama2-chat), it can provide better-than-random guess in terms of gene prediction.

However, the current LLM-based prediction did not match that of software specifically designed for this task. For instance, the best-performing GPT model demonstrated an average accuracy rate of 26.91% in the CSH dataset (24.31% for top 10 and 29.51% for top 50), which is notably better than random predictions (0.03% for top 10 and 0.17% for top 50). However, it falls significantly short of the accuracy achieved by Phen2Gene (35.3% for top 10 and 55.3% for top 50 with the same CSH dataset)[6]. Other methods, such as Phenolyzer and AMELIE (when using only HPO), also showed approximately 40% accuracy for the top 50 prediction experiments[39], which is still better than the best performance achieved by GPTs. Even worse, we have to keep in mind that the overall accuracy rate can be even lower because LLM cannot always guarantee to complete the task or complete it in the required output format. Furthermore, despite being capable of comprehending free-text input, LLMs, in general, achieved more accurate predictions when using human-curated HPO concepts as input rather than free-text inputs.

Additionally, we identify instances of hallucinations generated by all models, particularly when examining the top 50 predictions. These occurrences were often associated with generating of spurious gene variants within a gene family. For example, starting from OPA1, an actual existing gene, GPT-3.5 extends to OPA50 in sequence, which were not actual genes. Previous studies have highlighted the same limitations in answering medical questions[40,41] with attaching fabricated references. This deficiency may explain why GPT sometimes generates fictitious gene names when dealing with gene symbols, as gene symbols can function as both an acronym of the word as well as a type of identifier. One potential explanation is that Byte Pair Encoding (BPE) tokenizer, employed by both GPT and Llama2, breaks down the reference or gene symbols, allowing GPT the possibility to fabricate a reference or a gene symbol using its partial segments.

Bias was identified in the predicting specific genes. For example, when the actual diagnosed gene was 'TAF1', ‘DHX30’ and ‘WDR26’, none of the experiments produced accurate predictions. Conversely, when the diagnosed genes were 'MECP2', 'TP53' and ‘SCN1A’, GPT exhibited a considerably higher level of accuracy. For Llama2, ‘EP300’ and ‘WT1’ demonstrated high level of accuracy. A deeper investigation revealed that these genes have received substantial attention in research studies, with a higher number of Google Scholar search results, potentially contributing to their overprediction by LLMs. For genes those were not predicted accurately, the number of search hits was less than 10000 (e.g., TAF1: 9950, DXH30: 1010, WDR26: 972). In contrast, genes that were predicted with high accuracy had a significantly larger number of search hits (e.g., MECP2: 57700, TP53: 381000, SCN1A: 22300, EP300: 23000, WT1: 79300), except CDKL5 (8180), which had fewer cases than TAF1 but still exhibited high accuracy.

The above observations suggest a few important points. First, current LLMs might excel at tasks that humans can handle, but when it comes to tasks that humans are unable to manage and for which specialized knowledge bases were previously curated and models were trained upon specialized datasets, their performance does not match that of specialized tools. To enhance prediction accuracy in the future, fine-tuning the model with specialized datasets and models tailored to the genetic domain should be considered as in other task[42,43]. Studies have shown that in differential diagnosis tasks, the LLM model Med-PaLM2, which was fine-tuned with medical domain data based on the LLM model PaLM2[44], outperformed not only the regular GPT-4 in prediction accuracy by approximately 10% but also exceeded the performance of clinicians[45]. Second, we may still want to utilize HPO or other ontologies as intermediaries in the era of Large Language Models. Even though machines can now understand unstructured natural language, they still perform better when processing structured codes and concepts. Alternatively, prompts should be designed to break down the task further when the input is a natural language description, providing better guidance to GPT in achieving the prediction task through a two-step approach. Lastly, a general-purpose LLM is more susceptible to "common attention" bias. This is particularly significant in the context of rare disease diagnosis, as it suggests that general-purpose LLMs like GPTs may be more inclined to make predictions for commonly encountered cases but may not perform as effectively when dealing with rare conditions. Similar conclusions were drawn from the differential diagnosis tasks, where clinicians noted that LLMs were useful for simple cases. However, they had limitations for complex cases, which could not be assessed holistically but rather focused on specific aspects to draw conclusions[45].

Nevertheless, the future seems promising. When considering model size, we consistently observe a trend where larger models achieve better results across almost all evaluation metrics. The parameter size of GPT-3.5 and GPT-4 is approximately a hundred billion, which is larger than the 7 billion, 13 billion, and 70 billion parameter Llama2 models. While GPT-4 achieved the best performance in all three metrics, Llama2-70b-chat also outperformed its smaller counterparts in those metrics. This implies that in the future, even larger LLMs might have the ability to achieve performance comparable to, or even surpass, traditional knowledge graph-based bioinformatic tools. However, a plateau effect might also exist. Continuing evaluation and benchmark development are required to monitor the development of LLMs and their potential applications in the field of genomic medicine. considering GPT-4 is almost 35 times more expensive than GPT-3.5 (according to the 2010 August billing policy in OpenAI), crafting a more detailed prompt with more tokens for smaller LLMs might provide a more environmentally and economically friendly solution in certain tasks. Unfortunately, in our study, while we found different prompts could impact completeness and structural compliance in opposite ways, with no significant improvements in prediction accuracy were observed. Some previous studies have suggested the prompt influence was less dominant[46,47] for certain tasks. However, we believe that this is still a question that is not easily answered, especially considering that crafting an efficient prompt for LLMs remains an ongoing challenge. In the future, strategic model selection based on specific needs and constraints could be a significant research topic from both economic and computation speed perspectives. While the overall variation observed in the models' performance in this study is within acceptable parameters, the influence of the probabilistic nature of LLMs was still existed. Previous studies have suggested that the variability in results affected performance metrics when classifying functional evidence in biomedical literature[29]. Another study, involving a biomedical semantic question and answer task, which repeated five times with the same model, concluded that despite the presence of variability, its impact was minimal[30]. More investigation are needed to better understand how the variabilities of those LLMs could impact the downstream applications.

One limitation of our study is we exclusively focused on the zero-shot prompt scenario. Studies have shown that expanding GPT to include a more comprehensive range of examples and detailed steps could potentially lead to improvements in its performance. [48,49]. For example, transitioning from zero-shot to one-shot learning improved the relation extraction task by around 12%[50]. In another example, a 'chain-of-thoughts' prompt gave the best result for the summarization task compared to zero-shot performance[30].

## Conclusion

In this study, we conducted a comprehensive evaluation of the LLMs for phenotype-driven gene prioritization, a crucial process in rare genetic disorder diagnosis. Even the best-performing model, GPT-4, still lags behind traditional bioinformatics tools in terms of generating accurate candidate gene prediction results. However, a clear trend of LLM performance increasing with model size is observed. Notably, LLM's ability to process free-text phenotypic descriptions is advantageous, although it may not achieve the same level of robustness as terminology-based input. These findings contribute to the ongoing discussion about integrating advanced LLMs into clinical genomic analysis workflows.

## Acknowledgement

This study is supported by grant R01HG012655 from the National Human Genome Research Institute (NHGRI).

## Competing Interests

The authors declare no competing interests.

## Figure Legend

**Figure 1 Overall Design of the GPT Evaluation Study.** The prompt engineering process includes the combination of four different prompt templates, two types of phenotype representations, and requests for completing tasks at two difficulty levels. After engineering the prompts, they were input into both GPT-3.5 and GPT-4, each three times independently. The responses generated by GPT under different settings were collected for various outcome evaluations, including task completeness rate, gene prediction accuracy, and adherence to the output structure.

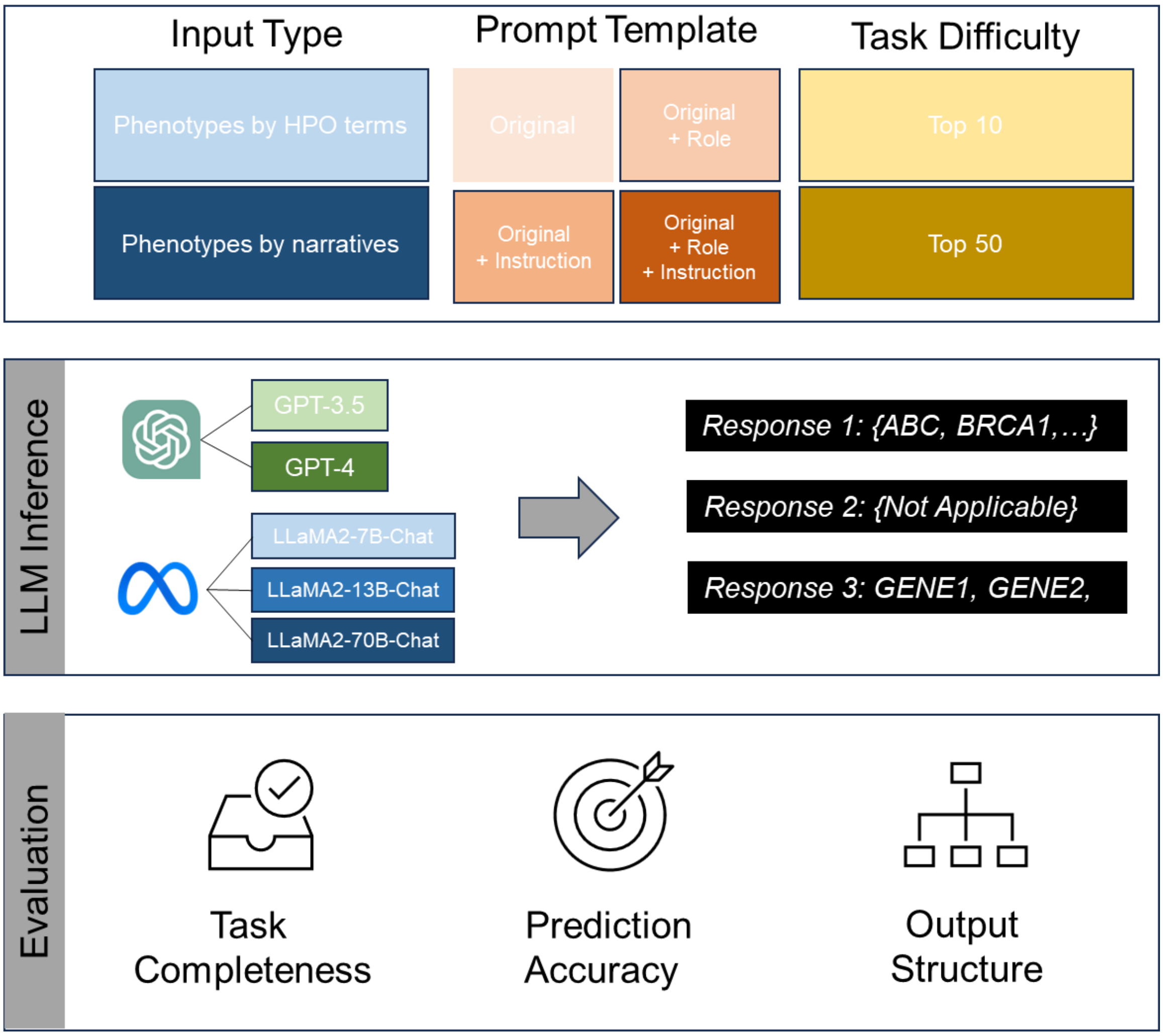

**Figure 2 Gene prediction accuracy across different models.** (A) Overall accuracy rate across different LLM models. (B) Comparison with other bioinformatics tools in predicting top 50 candidate genes in DGD datasets[6]; (C) Prediction accuracy across different datasets for GPT-4, specifically examining the top 50 gene prediction results.

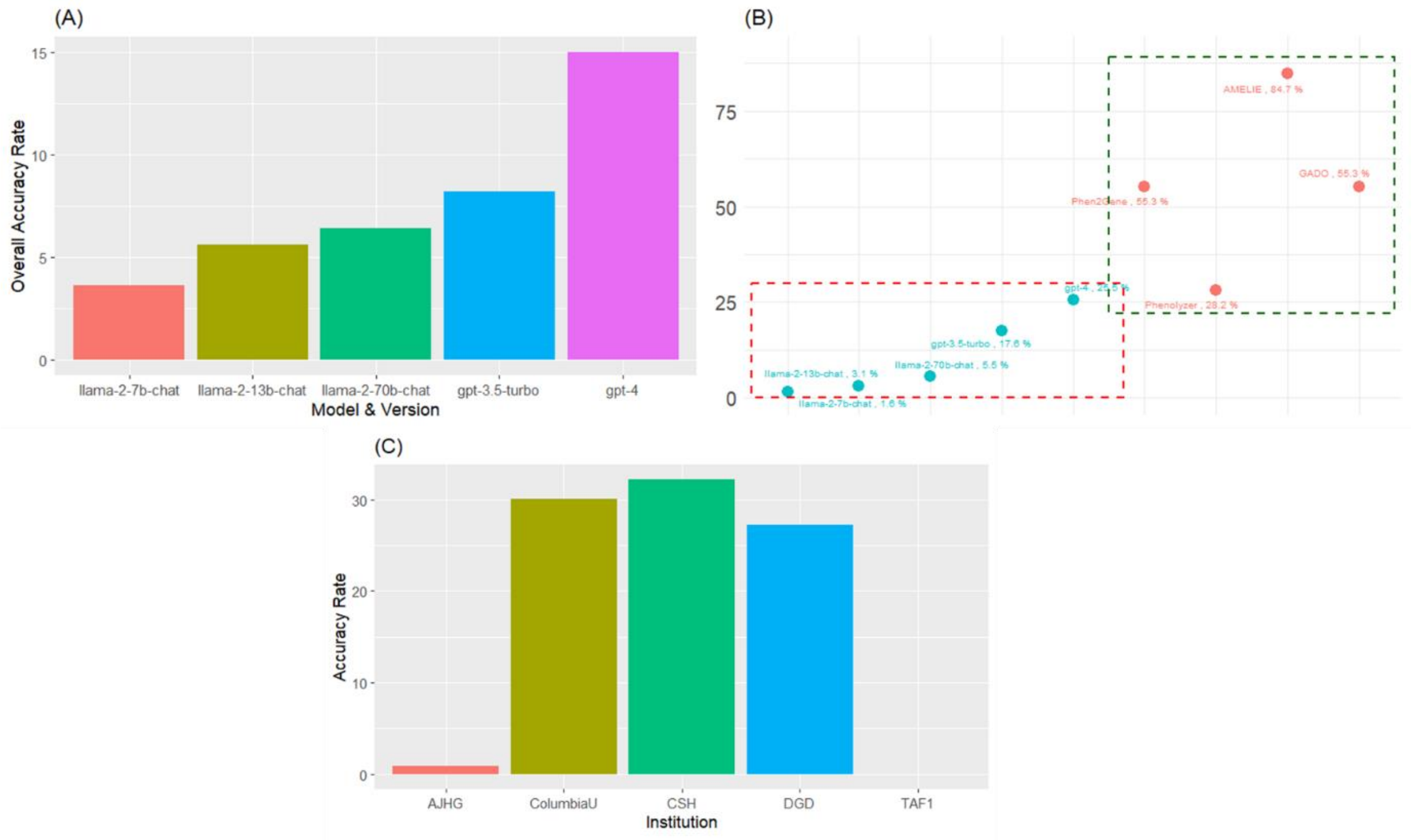

**Figure 3 Genes Most Frequently Predicted by LLMs Across All Experiments.** The green bar represents the number of times each gene was predicted in the challenging task (top 10), the blue bar shows the number of times each gene was predicted in the easier task (top 50), and the red bar indicates the number of times each gene appeared in the final diagnosed gene pool (i.e., ground truth occurrence). An asterisk (*) denotes genes that never appeared in the final diagnosed gene pool, indicating a high positive potential of LLMs for these genes.

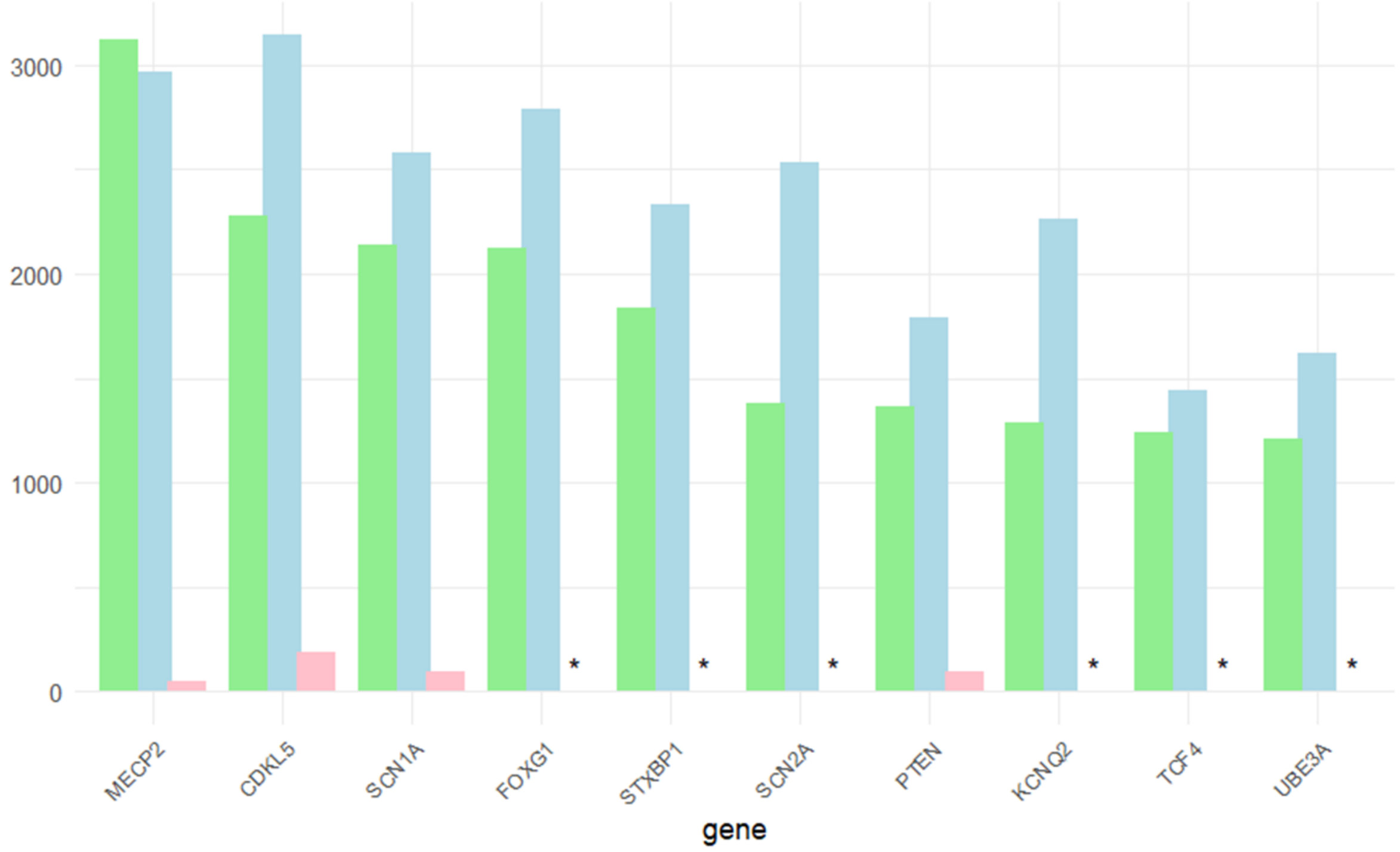

## Tables

**Table 1.** Dataset used for GPT evaluation.

| Source | # of individuals with HPO terms as input | # of individuals with narratives as input | Average # of HPO (s.d) | Average # of tokens (s.d) |
|---|---|---|---|---|
| AJHG | 78 | 72 | 11.42 (6.43) | 187.64 (88.87) |
| CSH | 72 | 49 | 12.83 (8.30) | 161.43 (102.38) |
| CUIMC | 27 | - | 11.52 (4.73) | - |
| DGD | 85 | - | 9.03 (4.06) | - |
| TAF1 | 14 | - | 34.86 (10.71) | - |
| Other | - | 4 | - | 180.25 (143.98) |

**Table 2.** Prompt engineering involved for phenotypic-based gene prioritization task. {Phenotype list} can be either a set of HPO-based concept names separated by ";" or a narrative containing phenotype descriptions extracted from the original literature. {top k} can be either "top 10" or "top 50," representing challenge and easy, respectively.

| Prompt template | Prompt content |
| --- | --- |
| Original | The phenotype description of the patient is {Phenotype list}. Can you suggest a list of {top k} possible genes to test? Please return gene symbols as a comma separated list. Example: "ABC1,BRAC2,BRAC1" or "Not Applicable" if you can not provide the result. . |
| Original + Role | Consider you are a genetic counselor. The phenotype description of the patient is {Phenotype list}. Can you suggest a list of {top k} possible genes to test? Please return gene symbols as a comma separated list. Example: "ABC1,BRAC2,BRAC1" or "Not Applicable" if you can not provide the result. |
| Original + Instruction | The phenotype description of the patient is {Phenotype list}. Can you suggest a list of {top k} possible genes to test? Please use the knowledge you have trained. No need to access to the real-time database to generate outcomes. Also, please return gene symbols as a comma separated list. Example: "ABC1,BRAC2,BRAC1"' or "Not Applicable" if you can not provide the result. |
| Original + Role + Instruction | Consider you are a genetic counselor. The phenotype description of the patient is {Phenotype list}. Can you suggest a list of {top k} possible genes to test? Please use the knowledge you have trained. There is no need to access the real-time database to generate outcomes. Also, please return gene symbols as a comma separated list. Example: "ABC1,BRAC2,BRAC1"' or "Not Applicable" if you can not provide the result. |

**Table 3. Evaluation results of the LLMs performance in making accuracy gene prediction for difficult tasks (Top 10). The number in the parenthesis is overall accuracy.**

| **Factor** | | **GPT-4** | **GPT-3.5** | **Llama2-7b-chat** | **Llama2-13b-chat** | **Llama2-70b-chat** |
|---|---|---|---|---|---|---|
| Prompts | Original | 13.83% (13.63%) | 7.20% (2.24%) | - | - | - |
| | Original + Role | 14.51% (14.13%) | 7.16% (4.49%) | - | - | - |
| | Original + Instruction | 14.17% (14.13%) | 10.45% (8.56%) | - | - | - |
| | Original + Role + Instruction | 13.93% (13.88%) | 10.21% (10.14%) | 5.34% (4.66%) | 7.15% (6.57%) | 7.46% (6.15%) |
| Input Type | HPO Concept | 15.06% (14.89%) | 9.53% (6.19%) | 5.41% (4.71%) | 7.41% (6.64%) | 8.21% (7.13%) |
| | Free Text | 12.00% (11.87%) | 8.71% (6.73%) | 5.20% (4.53%) | 6.61% (6.40%) | 5.49% (4.00%) |

**Table 4. Evaluation results of the LLMs performance in making accuracy gene prediction for easy tasks (Top 50). The number in the parenthesis is overall accuracy.**

| Factor | | GPT-4 | GPT-3.5 | Llama2-7b-chat | Llama2-13b-chat | Llama2-70b-chat |
|---|---|---|---|---|---|---|
| Prompts | Original | 17.87% (15.46%) | 15.93% (2.41%) | - | - | - |
| | Original + Role | 20.06% (16.38%) | 16.09% (10.89%) | - | - | - |
| | Original + Instruction | 16.35% (15.71%) | 16.04% (11.31%) | - | - | - |
| | Original + Role + Instruction | 18.11% (17.04%) | 17.28% (15.30%) | 4.75% (2.58%) | 8.66% (4.66%) | 7.99% (6.65%) |
| Input Type | HPO Concept | 20.67% (18.18%) | 17.87% (10.84%) | 5.39% (3.02%) | 10.88% (5.68%) | 9.09% (7.61%) |
| | Free Text | 12.25% (11.67%) | 13.44% (8.07%) | 3.17% (1.60%) | 4.19% (2.40%) | 5.52% (4.53%) |

**Table 5. Evaluation results of the LLMs performance in completing the task.**

| Factor | | GPT-4 | GPT-3.5 | Llama2-7b-chat | Llama2-13b-chat | Llama2-70b-chat |
|---|---|---|---|---|---|---|
| Prompts | Original | 92.56% | 23.15% | - | - | - |
| | Original + Role | 89.53% | 65.17% | - | - | - |
| | Original + Instruction | 97.92% | 76.23% | - | - | - |
| | Original + Role + Instruction | 96.88% | 93.93% | 70.70% | 72.82% | 82.83% |
| Task levels | Top 10 | 98.86% | 68.79% | 87.11% | 91.85% | 82.46% |
| | Top 50 | 89.59% | 60.45% | 54.28% | 53.78% | 83.21% |
| Input Type | HPO Concept | 93.40% | 62.79% | 71.56% | 70.89% | 85.27% |
| | Free Text | 96.03% | 68.67% | 68.80% | 77.07% | 77.47% |

**Table 6. Evaluation results of the LLMs performance in producing results adhering to the required output structure format.**

| Factor | | GPT-4 | GPT-3.5 | Llama2-7b-chat | Llama2-13b-chat | Llama2-70b-chat |
|---|---|---|---|---|---|---|
| Prompts | Original | 77.64% | 62.97% | - | - | - |
| | Original + Role | 80.71% | 30.59% | - | - | - |
| | Original + Instruction | 79.76% | 15.46% | - | - | - |
| | Original + Role + Instruction | 79.01% | 0.25% | 0.46% | 1.00% | 0.25% |
| Task levels | Top 10 | 99.98% | 31.17% | 0.08% | 0.25% | 0.17% |
| | Top 50 | 58.58% | 23.46% | 0.83% | 1.75% | 0.33% |
| Input Type | HPO Concept | 80.01% | 30.15% | 0.48% | 1.03% | 0.18% |
| | Free Text | 77.67% | 21.07% | 0.40% | 0.93% | 0.40% |

## Supplementary Material

**Table S1 Odds ratio calculated for the genes in the final diagnosed pools across different models.** The expected count was calculated as the probability of the gene appeared in the final diagnosed pools times the number of total predictions GPTs made, which is the summation of the number of genes predicted in each experiment. The observed count was calculated by counting the number of occurrences of the gene in the prediction.

**Table S2 Average performance and Standard Deviation across different factor combinations.** A detailed metric calculations subdivided by each factor combination, including LLM models, Input types, Prompts and Task difficulties. It showcases the variability and average performance metrics, providing a comprehensive overview of how each factor combination influences the model's efficacy.

**Supplementary Figure 1. Examples Illustrating Two Instances of GPT Responses Where the Task is Incomplete.** In the left panel, GPT declines to provide prediction results. In the right panel, GPT fabricates prediction genes as "Gene 1", "Gene 2", and "Gene 3".

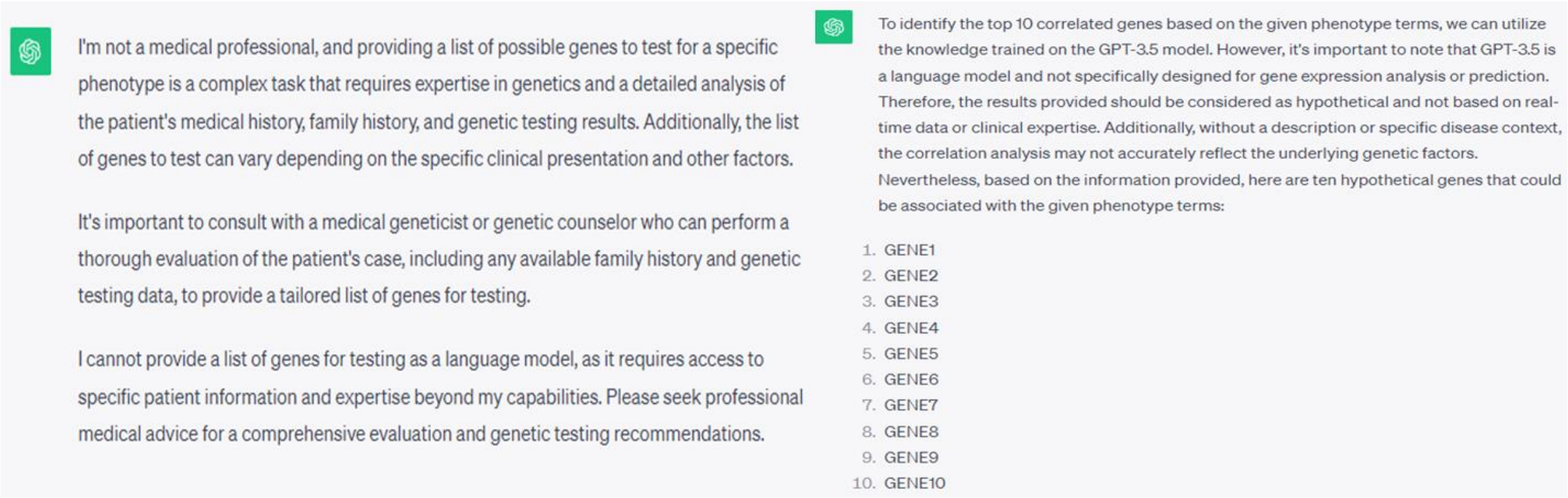

**Supplementary Figure 2. Examples of Two Instances of GPT Responses (Negative and Positive Examples) Indicating the Correct or Incorrect Gene Prediction for an Individual Whose Final Diagnosed Gene is 'ANK1.'** In the left panel, 'ANK1' is not present, indicating a wrong prediction. In the right panel, GPT provided a gene list containing 'ANK1,' indicating a correct prediction.

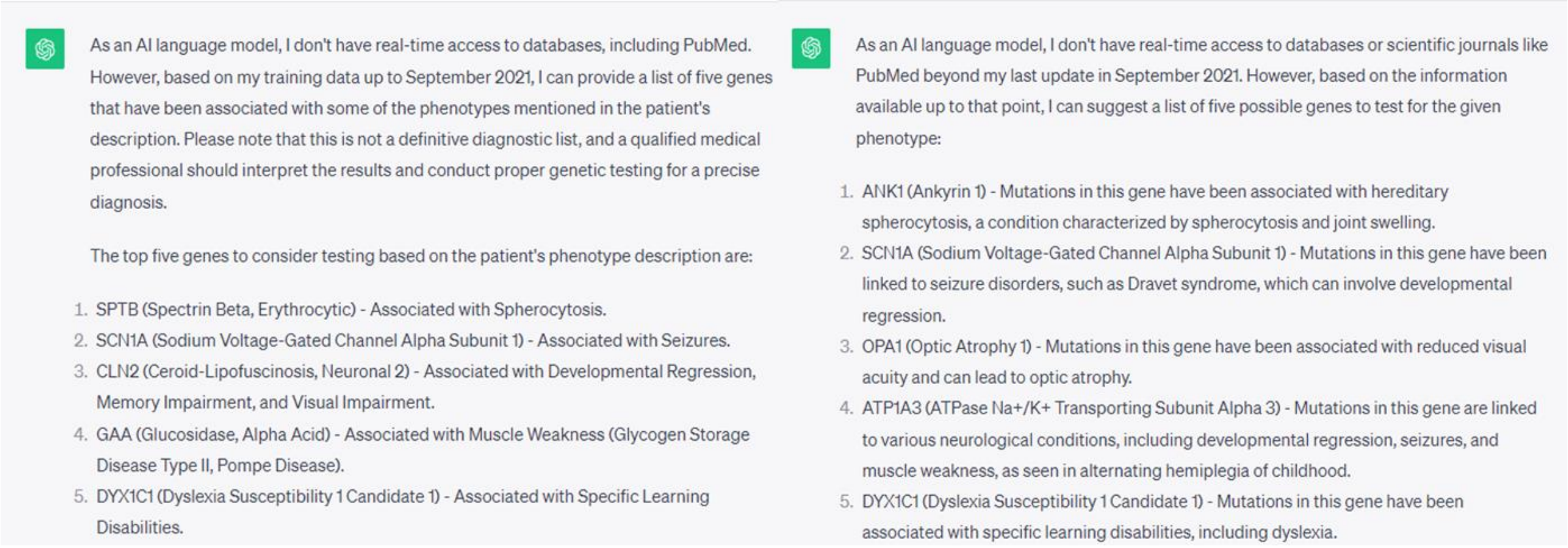

As an AI language model, I don't have real-time access to databases, including PubMed. However, based on my training data up to September 2021, I can provide a list of five genes that have been associated with some of the phenotypes mentioned in the patient's description. Please note that this is not a definitive diagnostic list, and a qualified medical professional should interpret the results and conduct proper genetic testing for a precise diagnosis.

The top five genes to consider testing based on the patient's phenotype description are:

1. SPTB (Spectrin Beta, Erythrocytic) - Associated with Spherocytosis.
2. SCN1A (Sodium Voltage-Gated Channel Alpha Subunit 1) - Associated with Seizures.
3. CLN2 (Ceroid-Lipofuscinosis, Neuronal 2) - Associated with Developmental Regression, Memory Impairment, and Visual Impairment.
4. GAA (Glucosidase, Alpha Acid) - Associated with Muscle Weakness (Glycogen Storage Disease Type II, Pompe Disease).
5. DYX1C1 (Dyslexia Susceptibility 1 Candidate 1) - Associated with Specific Learning Disabilities.

As an AI language model, I don't have real-time access to databases or scientific journals like PubMed beyond my last update in September 2021. However, based on the information available up to that point, I can suggest a list of five possible genes to test for the given phenotype:

1. ANK1 (Ankyrin 1) - Mutations in this gene have been associated with hereditary spherocytosis, a condition characterized by spherocytosis and joint swelling.
2. SCN1A (Sodium Voltage-Gated Channel Alpha Subunit 1) - Mutations in this gene have been linked to seizure disorders, such as Dravet syndrome, which can involve developmental regression.
3. OPA1 (Optic Atrophy 1) - Mutations in this gene have been associated with reduced visual acuity and can lead to optic atrophy.
4. ATP1A3 (ATPase Na+/K+ Transporting Subunit Alpha 3) - Mutations in this gene are linked to various neurological conditions, including developmental regression, seizures, and muscle weakness, as seen in alternating hemiplegia of childhood.
5. DYX1C1 (Dyslexia Susceptibility 1 Candidate 1) - Mutations in this gene have been associated with specific learning disabilities, including dyslexia.

**Supplementary Figure 3. Examples of Two Instances of GPT Responses That Do Not Adhere to the Required Output Structure. In the left panel, GPT did not provide a gene list nor 'Not Applicable' explicitly. In the right panel, although GPT generated a gene list, the format did not align with the required comma-separated gene list. A compliant result should consist of a gene list with the exact number of genes requested in the prompt, such as [OPA1, GJB2, GJB6, …], or generate 'Not Applicable'.**

As an AI language model, I can provide general information and insights, but I do not have access to real-time patient data or the ability to diagnose specific medical conditions. However, I can give you an example of how you might approach finding correlated genes based on phenotype terms. Please note that this is a hypothetical scenario and should not be used as a substitute for professional medical advice.

To identify correlated genes based on phenotype terms, you can utilize various bioinformatics and genomics tools that analyze gene expression data and phenotype associations. One commonly used approach is to perform gene set enrichment analysis (GSEA) or pathway analysis to identify gene sets that are significantly associated with the given phenotypes. These gene sets can then be used to identify the specific genes that are correlated with the phenotypes.

Without access to a real-time database, I can't provide you with a list of the top 50 correlated genes for the given phenotype terms. However, I can explain the general process you might follow to identify correlated genes:

1. Compile a list of genes associated with each phenotype term: Start by collecting a list of genes that have been previously associated with each phenotype term. This information can be obtained from various databases, literature reviews, or online resources that catalog gene-phenotype associations.
2. Combine the gene lists: Merge the gene lists obtained from each phenotype term to create a comprehensive list of genes associated with any of the given phenotypes. This will capture the genes that are potentially correlated with the overall phenotype presentation.
3. Perform gene set enrichment analysis: Utilize a bioinformatics tool or software that performs gene set enrichment analysis to identify gene sets or pathways that are significantly enriched among the compiled gene list. This analysis will help identify specific biological processes or pathways that are related to the given phenotypes.
4. Identify correlated genes: From the enriched gene sets or pathways, extract the individual genes that are contributing to the enrichment signal. These genes are likely to be correlated with the given phenotypes and may represent potential candidate genes involved in the observed phenotype presentation.

I apologize for the confusion, but as an AI language model, I don't have real-time access to databases or the ability to perform direct correlations between phenotype terms and genes. However, I can provide you with some general information about genes associated with the given phenotype terms. It's important to note that the relationship between phenotype terms and genes is complex and often involves multiple genes interacting with each other.

Here are some genes that have been associated with the provided phenotype terms:

1. OPA1 (Optic Atrophy 1)
2. GJB2 (Gap Junction Protein Beta 2)
3. GJB6 (Gap Junction Protein Beta 6)
4. SLC26A4 (Solute Carrier Family 26 Member 4)
5. POU3F4 (POU Class 3 Homeobox 4)
6. EYA1 (EYA Transcriptional Coactivator and Phosphatase 1)
7. SHOX (Short Stature Homeobox)
8. USH1C (Usher Syndrome 1C)
9. USH2A (Usher Syndrome 2A)
10. MYO7A (Myosin VIIA)
11. DFNB31 (Deafness, Autosomal Recessive 31)
12. DFNB59 (Deafness, Autosomal Recessive 59)
13. DFNA5 (Deafness, Autosomal Dominant 5)
14. GPR98 (G Protein-Coupled Receptor 98)
15. PDZD7 (PDZ Domain Containing 7)
16. STRC (Stereocilin)
17. TECTA (Tectorin Alpha)
18. TECTB (Tectorin Beta)
19. TJP2 (Tight Junction Protein 2)
20. TMIE (Transmembrane Inner Ear)

Please note that this is not an exhaustive list, and there could be additional genes associated with the provided phenotype terms. A comprehensive analysis would require more specific information about the patient, such as genetic testing results or a detailed medical evaluation. I would recommend consulting with a medical professional or geneticist who can provide a more accurate assessment based on the patient's individual case.

| true_gene | gpt_version | k_i | o_i | k | m | e_i | odds_ratio |
|---|---|---|---|---|---|---|---|
| BRCA1 | llama-2-7b-chat | 6 | 6 | 2406 | 87 | 0.216957605 | 27.65517241 |
| SHH | llama-2-7b-chat | 6 | 4 | 2406 | 87 | 0.216957605 | 18.43678160 |
| TP53 | llama-2-7b-chat | 84 | 46 | 2406 | 87 | 3.037406483 | 15.14449917 |
| ASXL1 | llama-2-7b-chat | 6 | 3 | 2406 | 87 | 0.216957605 | 13.82758620 |
| EP300 | llama-2-7b-chat | 6 | 3 | 2406 | 87 | 0.216957605 | 13.82758620 |
| LMNA | llama-2-7b-chat | 6 | 3 | 2406 | 87 | 0.216957605 | 13.82758620 |
| NPC1 | llama-2-7b-chat | 6 | 3 | 2406 | 87 | 0.216957605 | 13.82758620 |
| SDHA | llama-2-7b-chat | 6 | 3 | 2406 | 87 | 0.216957605 | 13.82758620 |
| SCN1A | llama-2-7b-chat | 12 | 3 | 2406 | 87 | 0.433915211 | 6.913793103 |
| CACNA1C | llama-2-7b-chat | 6 | 1 | 2406 | 87 | 0.216957605 | 4.609195402 |
| PTEN | llama-2-7b-chat | 12 | 2 | 2406 | 87 | 0.433915211 | 4.609195402 |
| SYNGAP1 | llama-2-7b-chat | 18 | 2 | 2406 | 87 | 0.650872817 | 3.072796934 |
| RAC1 | llama-2-7b-chat | 66 | 6 | 2406 | 87 | 2.386533665 | 2.514106583 |
| BRAF | llama-2-7b-chat | 24 | 2 | 2406 | 87 | 0.867830423 | 2.304597701 |
| ABCD1 | llama-2-7b-chat | 6 | 0 | 2406 | 87 | 0.216957605 | 0.00 |
| ACTG2 | llama-2-7b-chat | 6 | 0 | 2406 | 87 | 0.216957605 | 0.00 |
| ADA2 | llama-2-7b-chat | 6 | 0 | 2406 | 87 | 0.216957605 | 0.00 |
| ADSS1 | llama-2-7b-chat | 6 | 0 | 2406 | 87 | 0.216957605 | 0.00 |
| AGL | llama-2-7b-chat | 6 | 0 | 2406 | 87 | 0.216957605 | 0.00 |
| AHDC1 | llama-2-7b-chat | 18 | 0 | 2406 | 87 | 0.650872817 | 0.00 |
| AIFM1 | llama-2-7b-chat | 12 | 0 | 2406 | 87 | 0.433915211 | 0.00 |
| AIRE | llama-2-7b-chat | 12 | 0 | 2406 | 87 | 0.433915211 | 0.00 |
| ALG13 | llama-2-7b-chat | 6 | 0 | 2406 | 87 | 0.216957605 | 0.00 |
| ALPK3 | llama-2-7b-chat | 12 | 0 | 2406 | 87 | 0.433915211 | 0.00 |
| ANK1 | llama-2-7b-chat | 6 | 0 | 2406 | 87 | 0.216957605 | 0.00 |
| ANKRD11 | llama-2-7b-chat | 24 | 0 | 2406 | 87 | 0.867830423 | 0.00 |
| APTX | llama-2-7b-chat | 6 | 0 | 2406 | 87 | 0.216957605 | 0.00 |
| AQP4 | llama-2-7b-chat | 12 | 0 | 2406 | 87 | 0.433915211 | 0.00 |
| ARID1B | llama-2-7b-chat | 24 | 0 | 2406 | 87 | 0.867830423 | 0.00 |
| ARID2 | llama-2-7b-chat | 6 | 0 | 2406 | 87 | 0.216957605 | 0.00 |
| ARX | llama-2-7b-chat | 6 | 0 | 2406 | 87 | 0.216957605 | 0.00 |
| ASXL3 | llama-2-7b-chat | 18 | 0 | 2406 | 87 | 0.650872817 | 0.00 |
| ATN1 | llama-2-7b-chat | 6 | 0 | 2406 | 87 | 0.216957605 | 0.00 |
| ATP1A3 | llama-2-7b-chat | 18 | 0 | 2406 | 87 | 0.650872817 | 0.00 |
| ATRX | llama-2-7b-chat | 6 | 0 | 2406 | 87 | 0.216957605 | 0.00 |
| AUTS2 | llama-2-7b-chat | 6 | 0 | 2406 | 87 | 0.216957605 | 0.00 |
| B3GLCT | llama-2-7b-chat | 6 | 0 | 2406 | 87 | 0.216957605 | 0.00 |
| BICD2 | llama-2-7b-chat | 24 | 0 | 2406 | 87 | 0.867830423 | 0.00 |
| BTD | llama-2-7b-chat | 6 | 0 | 2406 | 87 | 0.216957605 | 0.00 |
| BTK | llama-2-7b-chat | 6 | 0 | 2406 | 87 | 0.216957605 | 0.00 |
| C3AR1 | llama-2-7b-chat | 12 | 0 | 2406 | 87 | 0.433915211 | 0.00 |
| CAMK4 | llama-2-7b-chat | 12 | 0 | 2406 | 87 | 0.433915211 | 0.00 |
| CAPN5 | llama-2-7b-chat | 12 | 0 | 2406 | 87 | 0.433915211 | 0.00 |
| CCN6 | llama-2-7b-chat | 6 | 0 | 2406 | 87 | 0.216957605 | 0.00 |
| CDK10 | llama-2-7b-chat | 96 | 0 | 2406 | 87 | 3.471321695 | 0.00 |
| CDK13 | llama-2-7b-chat | 6 | 0 | 2406 | 87 | 0.216957605 | 0.00 |
| CDKL5 | llama-2-7b-chat | 24 | 0 | 2406 | 87 | 0.867830423 | 0.00 |
| CHAMP1 | llama-2-7b-chat | 6 | 0 | 2406 | 87 | 0.216957605 | 0.00 |
| CLUAP1 | llama-2-7b-chat | 12 | 0 | 2406 | 87 | 0.433915211 | 0.00 |
| COL10A1 | llama-2-7b-chat | 6 | 0 | 2406 | 87 | 0.216957605 | 0.00 |
| COL18A1 | llama-2-7b-chat | 6 | 0 | 2406 | 87 | 0.216957605 | 0.00 |
| COL4A1 | llama-2-7b-chat | 6 | 0 | 2406 | 87 | 0.216957605 | 0.00 |
| COL7A1 | llama-2-7b-chat | 12 | 0 | 2406 | 87 | 0.433915211 | 0.00 |
| CTNNB1 | llama-2-7b-chat | 6 | 0 | 2406 | 87 | 0.216957605 | 0.00 |
| CTSA | llama-2-7b-chat | 6 | 0 | 2406 | 87 | 0.216957605 | 0.00 |
| CYP17A1 | llama-2-7b-chat | 6 | 0 | 2406 | 87 | 0.216957605 | 0.00 |
| DCAF17 | llama-2-7b-chat | 6 | 0 | 2406 | 87 | 0.216957605 | 0.00 |
| DDR2 | llama-2-7b-chat | 6 | 0 | 2406 | 87 | 0.216957605 | 0.00 |
| DDX11 | llama-2-7b-chat | 6 | 0 | 2406 | 87 | 0.216957605 | 0.00 |
| DDX3X | llama-2-7b-chat | 6 | 0 | 2406 | 87 | 0.216957605 | 0.00 |
| DHX30 | llama-2-7b-chat | 144 | 0 | 2406 | 87 | 5.206982543 | 0.00 |
| EBF3 | llama-2-7b-chat | 6 | 0 | 2406 | 87 | 0.216957605 | 0.00 |
| EFL1 | llama-2-7b-chat | 12 | 0 | 2406 | 87 | 0.433915211 | 0.00 |
| EGFR | llama-2-7b-chat | 12 | 0 | 2406 | 87 | 0.433915211 | 0.00 |
| EHMT1 | llama-2-7b-chat | 12 | 0 | 2406 | 87 | 0.433915211 | 0.00 |
| EIF2AK4 | llama-2-7b-chat | 6 | 0 | 2406 | 87 | 0.216957605 | 0.00 |
| F13A1 | llama-2-7b-chat | 6 | 0 | 2406 | 87 | 0.216957605 | 0.00 |
| F8 | llama-2-7b-chat | 6 | 0 | 2406 | 87 | 0.216957605 | 0.00 |
| FDXR | llama-2-7b-chat | 84 | 0 | 2406 | 87 | 3.037406483 | 0.00 |
| FGD1 | llama-2-7b-chat | 12 | 0 | 2406 | 87 | 0.433915211 | 0.00 |
| FOXP2 | llama-2-7b-chat | 12 | 0 | 2406 | 87 | 0.433915211 | 0.00 |
| G6PD | llama-2-7b-chat | 6 | 0 | 2406 | 87 | 0.216957605 | 0.00 |
| GABRA1 | llama-2-7b-chat | 6 | 0 | 2406 | 87 | 0.216957605 | 0.00 |
| GNAS | llama-2-7b-chat | 6 | 0 | 2406 | 87 | 0.216957605 | 0.00 |
| GRIN2B | llama-2-7b-chat | 6 | 0 | 2406 | 87 | 0.216957605 | 0.00 |
| GRM1 | llama-2-7b-chat | 24 | 0 | 2406 | 87 | 0.867830423 | 0.00 |
| HBB | llama-2-7b-chat | 6 | 0 | 2406 | 87 | 0.216957605 | 0.00 |
| HGF | llama-2-7b-chat | 18 | 0 | 2406 | 87 | 0.650872817 | 0.00 |
| HNF4A | llama-2-7b-chat | 6 | 0 | 2406 | 87 | 0.216957605 | 0.00 |
| HNRNPH2 | llama-2-7b-chat | 6 | 0 | 2406 | 87 | 0.216957605 | 0.00 |
| IFT172 | llama-2-7b-chat | 6 | 0 | 2406 | 87 | 0.216957605 | 0.00 |
| IL12RB1 | llama-2-7b-chat | 6 | 0 | 2406 | 87 | 0.216957605 | 0.00 |
| IQSEC2 | llama-2-7b-chat | 6 | 0 | 2406 | 87 | 0.216957605 | 0.00 |
| KAT6A | llama-2-7b-chat | 12 | 0 | 2406 | 87 | 0.433915211 | 0.00 |
| KAT6B | llama-2-7b-chat | 6 | 0 | 2406 | 87 | 0.216957605 | 0.00 |
| KCNK9 | llama-2-7b-chat | 6 | 0 | 2406 | 87 | 0.216957605 | 0.00 |
| KCNN4 | llama-2-7b-chat | 6 | 0 | 2406 | 87 | 0.216957605 | 0.00 |
| KMT2A | llama-2-7b-chat | 6 | 0 | 2406 | 87 | 0.216957605 | 0.00 |
| KMT2D | llama-2-7b-chat | 6 | 0 | 2406 | 87 | 0.216957605 | 0.00 |
| LIPT2 | llama-2-7b-chat | 36 | 0 | 2406 | 87 | 1.301745635 | 0.00 |
| MAF | llama-2-7b-chat | 6 | 0 | 2406 | 87 | 0.216957605 | 0.00 |
| MAN2B1 | llama-2-7b-chat | 6 | 0 | 2406 | 87 | 0.216957605 | 0.00 |
| MECP2 | llama-2-7b-chat | 6 | 0 | 2406 | 87 | 0.216957605 | 0.00 |
| MED13L | llama-2-7b-chat | 12 | 0 | 2406 | 87 | 0.433915211 | 0.00 |
| MEN1 | llama-2-7b-chat | 6 | 0 | 2406 | 87 | 0.216957605 | 0.00 |
| MMUT | llama-2-7b-chat | 6 | 0 | 2406 | 87 | 0.216957605 | 0.00 |
| MRPS34 | llama-2-7b-chat | 66 | 0 | 2406 | 87 | 2.386533665 | 0.00 |
| MYH10 | llama-2-7b-chat | 6 | 0 | 2406 | 87 | 0.216957605 | 0.00 |
| MYH9 | llama-2-7b-chat | 12 | 0 | 2406 | 87 | 0.433915211 | 0.00 |
| MYO7A | llama-2-7b-chat | 6 | 0 | 2406 | 87 | 0.216957605 | 0.00 |
| NAA15 | llama-2-7b-chat | 6 | 0 | 2406 | 87 | 0.216957605 | 0.00 |
| NPHS2 | llama-2-7b-chat | 6 | 0 | 2406 | 87 | 0.216957605 | 0.00 |
| NPR3 | llama-2-7b-chat | 30 | 0 | 2406 | 87 | 1.084788029 | 0.00 |
| NR2F1 | llama-2-7b-chat | 24 | 0 | 2406 | 87 | 0.867830423 | 0.00 |
| NSD1 | llama-2-7b-chat | 6 | 0 | 2406 | 87 | 0.216957605 | 0.00 |
| OPA3 | llama-2-7b-chat | 12 | 0 | 2406 | 87 | 0.433915211 | 0.00 |
| PACS1 | llama-2-7b-chat | 12 | 0 | 2406 | 87 | 0.433915211 | 0.00 |
| PAX6 | llama-2-7b-chat | 6 | 0 | 2406 | 87 | 0.216957605 | 0.00 |
| PEPD | llama-2-7b-chat | 6 | 0 | 2406 | 87 | 0.216957605 | 0.00 |
| PEX1 | llama-2-7b-chat | 6 | 0 | 2406 | 87 | 0.216957605 | 0.00 |
| PEX26 | llama-2-7b-chat | 6 | 0 | 2406 | 87 | 0.216957605 | 0.00 |
| PHIP | llama-2-7b-chat | 18 | 0 | 2406 | 87 | 0.650872817 | 0.00 |
| PIGN | llama-2-7b-chat | 6 | 0 | 2406 | 87 | 0.216957605 | 0.00 |
| PLA2G4A | llama-2-7b-chat | 6 | 0 | 2406 | 87 | 0.216957605 | 0.00 |
| PMM2 | llama-2-7b-chat | 12 | 0 | 2406 | 87 | 0.433915211 | 0.00 |
| PMPCA | llama-2-7b-chat | 18 | 0 | 2406 | 87 | 0.650872817 | 0.00 |
| POMT1 | llama-2-7b-chat | 6 | 0 | 2406 | 87 | 0.216957605 | 0.00 |
| PPP3CA | llama-2-7b-chat | 12 | 0 | 2406 | 87 | 0.433915211 | 0.00 |
| PRRT2 | llama-2-7b-chat | 12 | 0 | 2406 | 87 | 0.433915211 | 0.00 |
| PTPN11 | llama-2-7b-chat | 12 | 0 | 2406 | 87 | 0.433915211 | 0.00 |
| PUF60 | llama-2-7b-chat | 6 | 0 | 2406 | 87 | 0.216957605 | 0.00 |
| PURA | llama-2-7b-chat | 18 | 0 | 2406 | 87 | 0.650872817 | 0.00 |
| PYCR2 | llama-2-7b-chat | 6 | 0 | 2406 | 87 | 0.216957605 | 0.00 |
| RAB11B | llama-2-7b-chat | 60 | 0 | 2406 | 87 | 2.169576059 | 0.00 |
| REST | llama-2-7b-chat | 18 | 0 | 2406 | 87 | 0.650872817 | 0.00 |
| RET | llama-2-7b-chat | 6 | 0 | 2406 | 87 | 0.216957605 | 0.00 |
| RYR1 | llama-2-7b-chat | 6 | 0 | 2406 | 87 | 0.216957605 | 0.00 |
| SAMHD1 | llama-2-7b-chat | 12 | 0 | 2406 | 87 | 0.433915211 | 0.00 |
| SCN8A | llama-2-7b-chat | 12 | 0 | 2406 | 87 | 0.433915211 | 0.00 |
| SCNN1B | llama-2-7b-chat | 12 | 0 | 2406 | 87 | 0.433915211 | 0.00 |
| SLC12A2 | llama-2-7b-chat | 12 | 0 | 2406 | 87 | 0.433915211 | 0.00 |
| SLC19A3 | llama-2-7b-chat | 18 | 0 | 2406 | 87 | 0.650872817 | 0.00 |
| SLC1A4 | llama-2-7b-chat | 6 | 0 | 2406 | 87 | 0.216957605 | 0.00 |
| SMARCA2 | llama-2-7b-chat | 6 | 0 | 2406 | 87 | 0.216957605 | 0.00 |
| SMARCA4 | llama-2-7b-chat | 6 | 0 | 2406 | 87 | 0.216957605 | 0.00 |
| SMC1A | llama-2-7b-chat | 12 | 0 | 2406 | 87 | 0.433915211 | 0.00 |
| SNAP25 | llama-2-7b-chat | 6 | 0 | 2406 | 87 | 0.216957605 | 0.00 |
| SON | llama-2-7b-chat | 6 | 0 | 2406 | 87 | 0.216957605 | 0.00 |
| SOX9 | llama-2-7b-chat | 12 | 0 | 2406 | 87 | 0.433915211 | 0.00 |
| SPART | llama-2-7b-chat | 36 | 0 | 2406 | 87 | 1.301745635 | 0.00 |
| SPG11 | llama-2-7b-chat | 6 | 0 | 2406 | 87 | 0.216957605 | 0.00 |
| SPTA1 | llama-2-7b-chat | 6 | 0 | 2406 | 87 | 0.216957605 | 0.00 |
| STRC | llama-2-7b-chat | 6 | 0 | 2406 | 87 | 0.216957605 | 0.00 |
| SUFU | llama-2-7b-chat | 12 | 0 | 2406 | 87 | 0.433915211 | 0.00 |
| TAF1 | llama-2-7b-chat | 84 | 0 | 2406 | 87 | 3.037406483 | 0.00 |
| TBC1D23 | llama-2-7b-chat | 18 | 0 | 2406 | 87 | 0.650872817 | 0.00 |
| TBCK | llama-2-7b-chat | 6 | 0 | 2406 | 87 | 0.216957605 | 0.00 |
| TBX3 | llama-2-7b-chat | 6 | 0 | 2406 | 87 | 0.216957605 | 0.00 |
| TBXAS1 | llama-2-7b-chat | 6 | 0 | 2406 | 87 | 0.216957605 | 0.00 |
| TCOF1 | llama-2-7b-chat | 6 | 0 | 2406 | 87 | 0.216957605 | 0.00 |
| TMEM231 | llama-2-7b-chat | 6 | 0 | 2406 | 87 | 0.216957605 | 0.00 |
| TMEM87B | llama-2-7b-chat | 12 | 0 | 2406 | 87 | 0.433915211 | 0.00 |
| TSHR | llama-2-7b-chat | 6 | 0 | 2406 | 87 | 0.216957605 | 0.00 |
| TUBB3 | llama-2-7b-chat | 6 | 0 | 2406 | 87 | 0.216957605 | 0.00 |
| TWIST1 | llama-2-7b-chat | 6 | 0 | 2406 | 87 | 0.216957605 | 0.00 |
| TWNK | llama-2-7b-chat | 12 | 0 | 2406 | 87 | 0.433915211 | 0.00 |
| UNC80 | llama-2-7b-chat | 6 | 0 | 2406 | 87 | 0.216957605 | 0.00 |
| VARS1 | llama-2-7b-chat | 18 | 0 | 2406 | 87 | 0.650872817 | 0.00 |
| WDR26 | llama-2-7b-chat | 156 | 0 | 2406 | 87 | 5.640897755 | 0.00 |
| WDR45 | llama-2-7b-chat | 6 | 0 | 2406 | 87 | 0.216957605 | 0.00 |
| WFS1 | llama-2-7b-chat | 6 | 0 | 2406 | 87 | 0.216957605 | 0.00 |
| WT1 | llama-2-7b-chat | 6 | 0 | 2406 | 87 | 0.216957605 | 0.00 |
| WWOX | llama-2-7b-chat | 6 | 0 | 2406 | 87 | 0.216957605 | 0.00 |
| YWHAG | llama-2-7b-chat | 72 | 0 | 2406 | 87 | 2.603491271 | 0.00 |
| ZMYND11 | llama-2-7b-chat | 12 | 0 | 2406 | 87 | 0.433915211 | 0.00 |
| TP53 | llama-2-70b-chat | 84 | 68 | 2406 | 154 | 5.376558603 | 12.64749536 |
| PTEN | llama-2-70b-chat | 12 | 9 | 2406 | 154 | 0.768079800 | 11.71753246 |
| ASXL1 | llama-2-70b-chat | 6 | 4 | 2406 | 154 | 0.384039900 | 10.41558441 |
| BRCA1 | llama-2-70b-chat | 6 | 4 | 2406 | 154 | 0.384039900 | 10.41558441 |
| BTK | llama-2-70b-chat | 6 | 3 | 2406 | 154 | 0.384039900 | 7.81168831 |
| CDKL5 | llama-2-70b-chat | 24 | 12 | 2406 | 154 | 1.536159600 | 7.81168831 |
| MEN1 | llama-2-70b-chat | 6 | 3 | 2406 | 154 | 0.384039900 | 7.81168831 |
| MYO7A | llama-2-70b-chat | 6 | 3 | 2406 | 154 | 0.384039900 | 7.81168831 |
| SCN1A | llama-2-70b-chat | 12 | 6 | 2406 | 154 | 0.768079800 | 7.81168831 |
| SYNGAP1 | llama-2-70b-chat | 18 | 7 | 2406 | 154 | 1.152119700 | 6.07575757 |
| RET | llama-2-70b-chat | 6 | 2 | 2406 | 154 | 0.384039900 | 5.20779220 |
| RYR1 | llama-2-70b-chat | 6 | 2 | 2406 | 154 | 0.384039900 | 5.20779220 |
| TSHR | llama-2-70b-chat | 6 | 2 | 2406 | 154 | 0.384039900 | 5.20779220 |
| BRAF | llama-2-70b-chat | 24 | 6 | 2406 | 154 | 1.536159600 | 3.90584415 |
| PRRT2 | llama-2-70b-chat | 12 | 3 | 2406 | 154 | 0.768079800 | 3.90584415 |
| ACTG2 | llama-2-70b-chat | 6 | 1 | 2406 | 154 | 0.384039900 | 2.60389610 |
| CACNA1C | llama-2-70b-chat | 6 | 1 | 2406 | 154 | 0.384039900 | 2.60389610 |
| CCN6 | llama-2-70b-chat | 6 | 1 | 2406 | 154 | 0.384039900 | 2.60389610 |
| FOXP2 | llama-2-70b-chat | 12 | 2 | 2406 | 154 | 0.768079800 | 2.60389610 |
| HNF4A | llama-2-70b-chat | 6 | 1 | 2406 | 154 | 0.384039900 | 2.60389610 |
| KAT6A | llama-2-70b-chat | 12 | 2 | 2406 | 154 | 0.768079800 | 2.60389610 |
| LMNA | llama-2-70b-chat | 6 | 1 | 2406 | 154 | 0.384039900 | 2.60389610 |
| PAX6 | llama-2-70b-chat | 6 | 1 | 2406 | 154 | 0.384039900 | 2.60389610 |
| PTPN11 | llama-2-70b-chat | 12 | 2 | 2406 | 154 | 0.768079800 | 2.60389610 |
| SDHA | llama-2-70b-chat | 6 | 1 | 2406 | 154 | 0.384039900 | 2.603896103 |
| SHH | llama-2-70b-chat | 6 | 1 | 2406 | 154 | 0.384039900 | 2.603896103 |
| SOX9 | llama-2-70b-chat | 12 | 2 | 2406 | 154 | 0.768079800 | 2.603896103 |
| TCOF1 | llama-2-70b-chat | 6 | 1 | 2406 | 154 | 0.384039900 | 2.603896103 |
| EGFR | llama-2-70b-chat | 12 | 1 | 2406 | 154 | 0.768079800 | 1.301948051 |
| NR2F1 | llama-2-70b-chat | 24 | 2 | 2406 | 154 | 1.536159600 | 1.301948051 |
| ABCD1 | llama-2-70b-chat | 6 | 0 | 2406 | 154 | 0.384039900 | 0.00 |
| ADA2 | llama-2-70b-chat | 6 | 0 | 2406 | 154 | 0.384039900 | 0.00 |
| ADSS1 | llama-2-70b-chat | 6 | 0 | 2406 | 154 | 0.384039900 | 0.00 |
| AGL | llama-2-70b-chat | 6 | 0 | 2406 | 154 | 0.384039900 | 0.00 |
| AHDC1 | llama-2-70b-chat | 18 | 0 | 2406 | 154 | 1.152119700 | 0.00 |
| AIFM1 | llama-2-70b-chat | 12 | 0 | 2406 | 154 | 0.768079800 | 0.00 |
| AIRE | llama-2-70b-chat | 12 | 0 | 2406 | 154 | 0.768079800 | 0.00 |
| ALG13 | llama-2-70b-chat | 6 | 0 | 2406 | 154 | 0.384039900 | 0.00 |
| ALPK3 | llama-2-70b-chat | 12 | 0 | 2406 | 154 | 0.768079800 | 0.00 |
| ANK1 | llama-2-70b-chat | 6 | 0 | 2406 | 154 | 0.384039900 | 0.00 |
| ANKRD11 | llama-2-70b-chat | 24 | 0 | 2406 | 154 | 1.536159600 | 0.00 |
| APTX | llama-2-70b-chat | 6 | 0 | 2406 | 154 | 0.384039900 | 0.00 |
| AQP4 | llama-2-70b-chat | 12 | 0 | 2406 | 154 | 0.768079800 | 0.00 |
| ARID1B | llama-2-70b-chat | 24 | 0 | 2406 | 154 | 1.536159600 | 0.00 |
| ARID2 | llama-2-70b-chat | 6 | 0 | 2406 | 154 | 0.384039900 | 0.00 |
| ARX | llama-2-70b-chat | 6 | 0 | 2406 | 154 | 0.384039900 | 0.00 |
| ASXL3 | llama-2-70b-chat | 18 | 0 | 2406 | 154 | 1.152119700 | 0.00 |
| ATN1 | llama-2-70b-chat | 6 | 0 | 2406 | 154 | 0.384039900 | 0.00 |
| ATP1A3 | llama-2-70b-chat | 18 | 0 | 2406 | 154 | 1.152119700 | 0.00 |
| ATRX | llama-2-70b-chat | 6 | 0 | 2406 | 154 | 0.384039900 | 0.00 |
| AUTS2 | llama-2-70b-chat | 6 | 0 | 2406 | 154 | 0.384039900 | 0.00 |
| B3GLCT | llama-2-70b-chat | 6 | 0 | 2406 | 154 | 0.384039900 | 0.00 |
| BICD2 | llama-2-70b-chat | 24 | 0 | 2406 | 154 | 1.536159600 | 0.00 |
| BTD | llama-2-70b-chat | 6 | 0 | 2406 | 154 | 0.384039900 | 0.00 |
| C3AR1 | llama-2-70b-chat | 12 | 0 | 2406 | 154 | 0.768079800 | 0.00 |
| CAMK4 | llama-2-70b-chat | 12 | 0 | 2406 | 154 | 0.768079800 | 0.00 |
| CAPN5 | llama-2-70b-chat | 12 | 0 | 2406 | 154 | 0.768079800 | 0.00 |
| CDK10 | llama-2-70b-chat | 96 | 0 | 2406 | 154 | 6.144638403 | 0.00 |
| CDK13 | llama-2-70b-chat | 6 | 0 | 2406 | 154 | 0.384039900 | 0.00 |
| CHAMP1 | llama-2-70b-chat | 6 | 0 | 2406 | 154 | 0.384039900 | 0.00 |
| CLUAP1 | llama-2-70b-chat | 12 | 0 | 2406 | 154 | 0.768079800 | 0.00 |
| COL10A1 | llama-2-70b-chat | 6 | 0 | 2406 | 154 | 0.384039900 | 0.00 |
| COL18A1 | llama-2-70b-chat | 6 | 0 | 2406 | 154 | 0.384039900 | 0.00 |
| COL4A1 | llama-2-70b-chat | 6 | 0 | 2406 | 154 | 0.384039900 | 0.00 |
| COL7A1 | llama-2-70b-chat | 12 | 0 | 2406 | 154 | 0.768079800 | 0.00 |
| CTNNB1 | llama-2-70b-chat | 6 | 0 | 2406 | 154 | 0.384039900 | 0.00 |
| CTSA | llama-2-70b-chat | 6 | 0 | 2406 | 154 | 0.384039900 | 0.00 |
| CYP17A1 | llama-2-70b-chat | 6 | 0 | 2406 | 154 | 0.384039900 | 0.00 |
| DCAF17 | llama-2-70b-chat | 6 | 0 | 2406 | 154 | 0.384039900 | 0.00 |
| DDR2 | llama-2-70b-chat | 6 | 0 | 2406 | 154 | 0.384039900 | 0.00 |
| DDX11 | llama-2-70b-chat | 6 | 0 | 2406 | 154 | 0.384039900 | 0.00 |
| DDX3X | llama-2-70b-chat | 6 | 0 | 2406 | 154 | 0.384039900 | 0.00 |
| DHX30 | llama-2-70b-chat | 144 | 0 | 2406 | 154 | 9.216957605 | 0.00 |
| EBF3 | llama-2-70b-chat | 6 | 0 | 2406 | 154 | 0.384039900 | 0.00 |
| EFL1 | llama-2-70b-chat | 12 | 0 | 2406 | 154 | 0.768079800 | 0.00 |
| EHMT1 | llama-2-70b-chat | 12 | 0 | 2406 | 154 | 0.768079800 | 0.00 |
| EIF2AK4 | llama-2-70b-chat | 6 | 0 | 2406 | 154 | 0.384039900 | 0.00 |
| EP300 | llama-2-70b-chat | 6 | 0 | 2406 | 154 | 0.384039900 | 0.00 |
| F13A1 | llama-2-70b-chat | 6 | 0 | 2406 | 154 | 0.384039900 | 0.00 |
| F8 | llama-2-70b-chat | 6 | 0 | 2406 | 154 | 0.384039900 | 0.00 |
| FDXR | llama-2-70b-chat | 84 | 0 | 2406 | 154 | 5.376558603 | 0.00 |
| FGD1 | llama-2-70b-chat | 12 | 0 | 2406 | 154 | 0.768079800 | 0.00 |
| G6PD | llama-2-70b-chat | 6 | 0 | 2406 | 154 | 0.384039900 | 0.00 |
| GABRA1 | llama-2-70b-chat | 6 | 0 | 2406 | 154 | 0.384039900 | 0.00 |
| GNAS | llama-2-70b-chat | 6 | 0 | 2406 | 154 | 0.384039900 | 0.00 |
| GRIN2B | llama-2-70b-chat | 6 | 0 | 2406 | 154 | 0.384039900 | 0.00 |
| GRM1 | llama-2-70b-chat | 24 | 0 | 2406 | 154 | 1.536159600 | 0.00 |
| HBB | llama-2-70b-chat | 6 | 0 | 2406 | 154 | 0.384039900 | 0.00 |
| HGF | llama-2-70b-chat | 18 | 0 | 2406 | 154 | 1.152119700 | 0.00 |
| HNRNPH2 | llama-2-70b-chat | 6 | 0 | 2406 | 154 | 0.384039900 | 0.00 |
| IFT172 | llama-2-70b-chat | 6 | 0 | 2406 | 154 | 0.384039900 | 0.00 |
| IL12RB1 | llama-2-70b-chat | 6 | 0 | 2406 | 154 | 0.384039900 | 0.00 |
| IQSEC2 | llama-2-70b-chat | 6 | 0 | 2406 | 154 | 0.384039900 | 0.00 |
| KAT6B | llama-2-70b-chat | 6 | 0 | 2406 | 154 | 0.384039900 | 0.00 |
| KCNK9 | llama-2-70b-chat | 6 | 0 | 2406 | 154 | 0.384039900 | 0.00 |
| KCNN4 | llama-2-70b-chat | 6 | 0 | 2406 | 154 | 0.384039900 | 0.00 |
| KMT2A | llama-2-70b-chat | 6 | 0 | 2406 | 154 | 0.384039900 | 0.00 |
| KMT2D | llama-2-70b-chat | 6 | 0 | 2406 | 154 | 0.384039900 | 0.00 |
| LIPT2 | llama-2-70b-chat | 36 | 0 | 2406 | 154 | 2.304239401 | 0.00 |
| MAF | llama-2-70b-chat | 6 | 0 | 2406 | 154 | 0.384039900 | 0.00 |
| MAN2B1 | llama-2-70b-chat | 6 | 0 | 2406 | 154 | 0.384039900 | 0.00 |
| MECP2 | llama-2-70b-chat | 6 | 0 | 2406 | 154 | 0.384039900 | 0.00 |
| MED13L | llama-2-70b-chat | 12 | 0 | 2406 | 154 | 0.768079800 | 0.00 |
| MMUT | llama-2-70b-chat | 6 | 0 | 2406 | 154 | 0.384039900 | 0.00 |
| MRPS34 | llama-2-70b-chat | 66 | 0 | 2406 | 154 | 4.224438902 | 0.00 |
| MYH10 | llama-2-70b-chat | 6 | 0 | 2406 | 154 | 0.384039900 | 0.00 |
| MYH9 | llama-2-70b-chat | 12 | 0 | 2406 | 154 | 0.768079800 | 0.00 |
| NAA15 | llama-2-70b-chat | 6 | 0 | 2406 | 154 | 0.384039900 | 0.00 |
| NPC1 | llama-2-70b-chat | 6 | 0 | 2406 | 154 | 0.384039900 | 0.00 |
| NPHS2 | llama-2-70b-chat | 6 | 0 | 2406 | 154 | 0.384039900 | 0.00 |
| NPR3 | llama-2-70b-chat | 30 | 0 | 2406 | 154 | 1.920199501 | 0.00 |
| NSD1 | llama-2-70b-chat | 6 | 0 | 2406 | 154 | 0.384039900 | 0.00 |
| OPA3 | llama-2-70b-chat | 12 | 0 | 2406 | 154 | 0.768079800 | 0.00 |
| PACS1 | llama-2-70b-chat | 12 | 0 | 2406 | 154 | 0.768079800 | 0.00 |
| PEPD | llama-2-70b-chat | 6 | 0 | 2406 | 154 | 0.384039900 | 0.00 |
| PEX1 | llama-2-70b-chat | 6 | 0 | 2406 | 154 | 0.384039900 | 0.00 |
| PEX26 | llama-2-70b-chat | 6 | 0 | 2406 | 154 | 0.384039900 | 0.00 |
| PHIP | llama-2-70b-chat | 18 | 0 | 2406 | 154 | 1.152119700 | 0.00 |
| PIGN | llama-2-70b-chat | 6 | 0 | 2406 | 154 | 0.384039900 | 0.00 |
| PLA2G4A | llama-2-70b-chat | 6 | 0 | 2406 | 154 | 0.384039900 | 0.00 |
| PMM2 | llama-2-70b-chat | 12 | 0 | 2406 | 154 | 0.768079800 | 0.00 |
| PMPCA | llama-2-70b-chat | 18 | 0 | 2406 | 154 | 1.152119700 | 0.00 |
| POMT1 | llama-2-70b-chat | 6 | 0 | 2406 | 154 | 0.384039900 | 0.00 |
| PPP3CA | llama-2-70b-chat | 12 | 0 | 2406 | 154 | 0.768079800 | 0.00 |
| PUF60 | llama-2-70b-chat | 6 | 0 | 2406 | 154 | 0.384039900 | 0.00 |
| PURA | llama-2-70b-chat | 18 | 0 | 2406 | 154 | 1.152119700 | 0.00 |
| PYCR2 | llama-2-70b-chat | 6 | 0 | 2406 | 154 | 0.384039900 | 0.00 |
| RAB11B | llama-2-70b-chat | 60 | 0 | 2406 | 154 | 3.840399002 | 0.00 |
| RAC1 | llama-2-70b-chat | 66 | 0 | 2406 | 154 | 4.224438902 | 0.00 |
| REST | llama-2-70b-chat | 18 | 0 | 2406 | 154 | 1.152119700 | 0.00 |
| SAMHD1 | llama-2-70b-chat | 12 | 0 | 2406 | 154 | 0.768079800 | 0.00 |
| SCN8A | llama-2-70b-chat | 12 | 0 | 2406 | 154 | 0.768079800 | 0.00 |
| SCNN1B | llama-2-70b-chat | 12 | 0 | 2406 | 154 | 0.768079800 | 0.00 |
| SLC12A2 | llama-2-70b-chat | 12 | 0 | 2406 | 154 | 0.768079800 | 0.00 |
| SLC19A3 | llama-2-70b-chat | 18 | 0 | 2406 | 154 | 1.152119700 | 0.00 |
| SLC1A4 | llama-2-70b-chat | 6 | 0 | 2406 | 154 | 0.384039900 | 0.00 |
| SMARCA2 | llama-2-70b-chat | 6 | 0 | 2406 | 154 | 0.384039900 | 0.00 |
| SMARCA4 | llama-2-70b-chat | 6 | 0 | 2406 | 154 | 0.384039900 | 0.00 |
| SMC1A | llama-2-70b-chat | 12 | 0 | 2406 | 154 | 0.768079800 | 0.00 |
| SNAP25 | llama-2-70b-chat | 6 | 0 | 2406 | 154 | 0.384039900 | 0.00 |
| SON | llama-2-70b-chat | 6 | 0 | 2406 | 154 | 0.384039900 | 0.00 |
| SPART | llama-2-70b-chat | 36 | 0 | 2406 | 154 | 2.304239401 | 0.00 |
| SPG11 | llama-2-70b-chat | 6 | 0 | 2406 | 154 | 0.384039900 | 0.00 |
| SPTA1 | llama-2-70b-chat | 6 | 0 | 2406 | 154 | 0.384039900 | 0.00 |
| STRC | llama-2-70b-chat | 6 | 0 | 2406 | 154 | 0.384039900 | 0.00 |
| SUFU | llama-2-70b-chat | 12 | 0 | 2406 | 154 | 0.768079800 | 0.00 |
| TAF1 | llama-2-70b-chat | 84 | 0 | 2406 | 154 | 5.376558603 | 0.00 |
| TBC1D23 | llama-2-70b-chat | 18 | 0 | 2406 | 154 | 1.152119700 | 0.00 |
| TBCK | llama-2-70b-chat | 6 | 0 | 2406 | 154 | 0.384039900 | 0.00 |
| TBX3 | llama-2-70b-chat | 6 | 0 | 2406 | 154 | 0.384039900 | 0.00 |
| TBXAS1 | llama-2-70b-chat | 6 | 0 | 2406 | 154 | 0.384039900 | 0.00 |
| TMEM231 | llama-2-70b-chat | 6 | 0 | 2406 | 154 | 0.384039900 | 0.00 |
| TMEM87B | llama-2-70b-chat | 12 | 0 | 2406 | 154 | 0.768079800 | 0.00 |
| TUBB3 | llama-2-70b-chat | 6 | 0 | 2406 | 154 | 0.384039900 | 0.00 |
| TWIST1 | llama-2-70b-chat | 6 | 0 | 2406 | 154 | 0.384039900 | 0.00 |
| TWNK | llama-2-70b-chat | 12 | 0 | 2406 | 154 | 0.768079800 | 0.00 |
| UNC80 | llama-2-70b-chat | 6 | 0 | 2406 | 154 | 0.384039900 | 0.00 |
| VARS1 | llama-2-70b-chat | 18 | 0 | 2406 | 154 | 1.152119700 | 0.00 |
| WDR26 | llama-2-70b-chat | 156 | 0 | 2406 | 154 | 9.99 | 0.00 |
| WDR45 | llama-2-70b-chat | 6 | 0 | 2406 | 154 | 0.384039900 | 0.00 |
| WFS1 | llama-2-70b-chat | 6 | 0 | 2406 | 154 | 0.384039900 | 0.00 |
| WT1 | llama-2-70b-chat | 6 | 0 | 2406 | 154 | 0.384039900 | 0.00 |
| WWOX | llama-2-70b-chat | 6 | 0 | 2406 | 154 | 0.384039900 | 0.00 |
| YWHAG | llama-2-70b-chat | 72 | 0 | 2406 | 154 | 4.608478802 | 0.00 |
| ZMYND11 | llama-2-70b-chat | 12 | 0 | 2406 | 154 | 0.768079800 | 0.00 |
| LMNA | llama-2-13b-chat | 6 | 6 | 2406 | 135 | 0.336658354 | 17.8222222 |
| PTEN | llama-2-13b-chat | 12 | 10 | 2406 | 135 | 0.673316708 | 14.8518518 |
| TP53 | llama-2-13b-chat | 84 | 62 | 2406 | 135 | 4.713216957 | 13.1544973 |
| BRCA1 | llama-2-13b-chat | 6 | 3 | 2406 | 135 | 0.336658354 | 8.91111111 |
| WT1 | llama-2-13b-chat | 6 | 3 | 2406 | 135 | 0.336658354 | 8.91111111 |
| BRAF | llama-2-13b-chat | 24 | 8 | 2406 | 135 | 1.346633416 | 5.94074074 |
| CTNNB1 | llama-2-13b-chat | 6 | 2 | 2406 | 135 | 0.336658354 | 5.94074074 |
| PTPN11 | llama-2-13b-chat | 12 | 4 | 2406 | 135 | 0.673316708 | 5.94074074 |
| SPG11 | llama-2-13b-chat | 6 | 2 | 2406 | 135 | 0.336658354 | 5.94074074 |
| TSHR | llama-2-13b-chat | 6 | 2 | 2406 | 135 | 0.336658354 | 5.94074074 |
| CDKL5 | llama-2-13b-chat | 24 | 7 | 2406 | 135 | 1.346633416 | 5.19814814 |
| AIRE | llama-2-13b-chat | 12 | 3 | 2406 | 135 | 0.673316708 | 4.455555555 |
| EGFR | llama-2-13b-chat | 12 | 3 | 2406 | 135 | 0.673316708 | 4.455555555 |
| SCN1A | llama-2-13b-chat | 12 | 3 | 2406 | 135 | 0.673316708 | 4.455555555 |
| SOX9 | llama-2-13b-chat | 12 | 3 | 2406 | 135 | 0.673316708 | 4.455555555 |
| IL12RB1 | llama-2-13b-chat | 6 | 1 | 2406 | 135 | 0.336658354 | 2.970370370 |
| MYO7A | llama-2-13b-chat | 6 | 1 | 2406 | 135 | 0.336658354 | 2.970370370 |
| RYR1 | llama-2-13b-chat | 6 | 1 | 2406 | 135 | 0.336658354 | 2.970370370 |
| SMARCA4 | llama-2-13b-chat | 6 | 1 | 2406 | 135 | 0.336658354 | 2.970370370 |
| SYNGAP1 | llama-2-13b-chat | 18 | 3 | 2406 | 135 | 1.009975062 | 2.970370370 |
| WDR45 | llama-2-13b-chat | 6 | 1 | 2406 | 135 | 0.336658354 | 2.970370370 |
| GRM1 | llama-2-13b-chat | 24 | 2 | 2406 | 135 | 1.346633416 | 1.485185185 |
| NR2F1 | llama-2-13b-chat | 24 | 2 | 2406 | 135 | 1.346633416 | 1.485185185 |
| PMM2 | llama-2-13b-chat | 12 | 1 | 2406 | 135 | 0.673316708 | 1.485185185 |
| SCN8A | llama-2-13b-chat | 12 | 1 | 2406 | 135 | 0.673316708 | 1.485185185 |
| ABCD1 | llama-2-13b-chat | 6 | 0 | 2406 | 135 | 0.336658354 | 0.00 |
| ACTG2 | llama-2-13b-chat | 6 | 0 | 2406 | 135 | 0.336658354 | 0.00 |
| ADA2 | llama-2-13b-chat | 6 | 0 | 2406 | 135 | 0.336658354 | 0.00 |
| ADSS1 | llama-2-13b-chat | 6 | 0 | 2406 | 135 | 0.336658354 | 0.00 |
| AGL | llama-2-13b-chat | 6 | 0 | 2406 | 135 | 0.336658354 | 0.00 |
| AHDC1 | llama-2-13b-chat | 18 | 0 | 2406 | 135 | 1.009975062 | 0.00 |
| AIFM1 | llama-2-13b-chat | 12 | 0 | 2406 | 135 | 0.673316708 | 0.00 |
| ALG13 | llama-2-13b-chat | 6 | 0 | 2406 | 135 | 0.336658354 | 0.00 |
| ALPK3 | llama-2-13b-chat | 12 | 0 | 2406 | 135 | 0.673316708 | 0.00 |
| ANK1 | llama-2-13b-chat | 6 | 0 | 2406 | 135 | 0.336658354 | 0.00 |
| ANKRD11 | llama-2-13b-chat | 24 | 0 | 2406 | 135 | 1.346633416 | 0.00 |
| APTX | llama-2-13b-chat | 6 | 0 | 2406 | 135 | 0.336658354 | 0.00 |
| AQP4 | llama-2-13b-chat | 12 | 0 | 2406 | 135 | 0.673316708 | 0.00 |
| ARID1B | llama-2-13b-chat | 24 | 0 | 2406 | 135 | 1.346633416 | 0.00 |
| ARID2 | llama-2-13b-chat | 6 | 0 | 2406 | 135 | 0.336658354 | 0.00 |
| ARX | llama-2-13b-chat | 6 | 0 | 2406 | 135 | 0.336658354 | 0.00 |
| ASXL1 | llama-2-13b-chat | 6 | 0 | 2406 | 135 | 0.336658354 | 0.00 |
| ASXL3 | llama-2-13b-chat | 18 | 0 | 2406 | 135 | 1.009975062 | 0.00 |
| ATN1 | llama-2-13b-chat | 6 | 0 | 2406 | 135 | 0.336658354 | 0.00 |
| ATP1A3 | llama-2-13b-chat | 18 | 0 | 2406 | 135 | 1.009975062 | 0.00 |
| ATRX | llama-2-13b-chat | 6 | 0 | 2406 | 135 | 0.336658354 | 0.00 |
| AUTS2 | llama-2-13b-chat | 6 | 0 | 2406 | 135 | 0.336658354 | 0.00 |
| B3GLCT | llama-2-13b-chat | 6 | 0 | 2406 | 135 | 0.336658354 | 0.00 |
| BICD2 | llama-2-13b-chat | 24 | 0 | 2406 | 135 | 1.346633416 | 0.00 |
| BTD | llama-2-13b-chat | 6 | 0 | 2406 | 135 | 0.336658354 | 0.00 |
| BTK | llama-2-13b-chat | 6 | 0 | 2406 | 135 | 0.336658354 | 0.00 |
| C3AR1 | llama-2-13b-chat | 12 | 0 | 2406 | 135 | 0.673316708 | 0.00 |
| CACNA1C | llama-2-13b-chat | 6 | 0 | 2406 | 135 | 0.336658354 | 0.00 |
| CAMK4 | llama-2-13b-chat | 12 | 0 | 2406 | 135 | 0.673316708 | 0.00 |
| CAPN5 | llama-2-13b-chat | 12 | 0 | 2406 | 135 | 0.673316708 | 0.00 |
| CCN6 | llama-2-13b-chat | 6 | 0 | 2406 | 135 | 0.336658354 | 0.00 |
| CDK10 | llama-2-13b-chat | 96 | 0 | 2406 | 135 | 5.386533665 | 0.00 |
| CDK13 | llama-2-13b-chat | 6 | 0 | 2406 | 135 | 0.336658354 | 0.00 |
| CHAMP1 | llama-2-13b-chat | 6 | 0 | 2406 | 135 | 0.336658354 | 0.00 |
| CLUAP1 | llama-2-13b-chat | 12 | 0 | 2406 | 135 | 0.673316708 | 0.00 |
| COL10A1 | llama-2-13b-chat | 6 | 0 | 2406 | 135 | 0.336658354 | 0.00 |
| COL18A1 | llama-2-13b-chat | 6 | 0 | 2406 | 135 | 0.336658354 | 0.00 |
| COL4A1 | llama-2-13b-chat | 6 | 0 | 2406 | 135 | 0.336658354 | 0.00 |
| COL7A1 | llama-2-13b-chat | 12 | 0 | 2406 | 135 | 0.673316708 | 0.00 |
| CTSA | llama-2-13b-chat | 6 | 0 | 2406 | 135 | 0.336658354 | 0.00 |
| CYP17A1 | llama-2-13b-chat | 6 | 0 | 2406 | 135 | 0.336658354 | 0.00 |
| DCAF17 | llama-2-13b-chat | 6 | 0 | 2406 | 135 | 0.336658354 | 0.00 |
| DDR2 | llama-2-13b-chat | 6 | 0 | 2406 | 135 | 0.336658354 | 0.00 |
| DDX11 | llama-2-13b-chat | 6 | 0 | 2406 | 135 | 0.336658354 | 0.00 |
| DDX3X | llama-2-13b-chat | 6 | 0 | 2406 | 135 | 0.336658354 | 0.00 |
| DHX30 | llama-2-13b-chat | 144 | 0 | 2406 | 135 | 8.079800498 | 0.00 |
| EBF3 | llama-2-13b-chat | 6 | 0 | 2406 | 135 | 0.336658354 | 0.00 |
| EFL1 | llama-2-13b-chat | 12 | 0 | 2406 | 135 | 0.673316708 | 0.00 |
| EHMT1 | llama-2-13b-chat | 12 | 0 | 2406 | 135 | 0.673316708 | 0.00 |
| EIF2AK4 | llama-2-13b-chat | 6 | 0 | 2406 | 135 | 0.336658354 | 0.00 |
| EP300 | llama-2-13b-chat | 6 | 0 | 2406 | 135 | 0.336658354 | 0.00 |
| F13A1 | llama-2-13b-chat | 6 | 0 | 2406 | 135 | 0.336658354 | 0.00 |
| F8 | llama-2-13b-chat | 6 | 0 | 2406 | 135 | 0.336658354 | 0.00 |
| FDXR | llama-2-13b-chat | 84 | 0 | 2406 | 135 | 4.713216957 | 0.00 |
| FGD1 | llama-2-13b-chat | 12 | 0 | 2406 | 135 | 0.673316708 | 0.00 |
| FOXP2 | llama-2-13b-chat | 12 | 0 | 2406 | 135 | 0.673316708 | 0.00 |
| G6PD | llama-2-13b-chat | 6 | 0 | 2406 | 135 | 0.336658354 | 0.00 |
| GABRA1 | llama-2-13b-chat | 6 | 0 | 2406 | 135 | 0.336658354 | 0.00 |
| GNAS | llama-2-13b-chat | 6 | 0 | 2406 | 135 | 0.336658354 | 0.00 |
| GRIN2B | llama-2-13b-chat | 6 | 0 | 2406 | 135 | 0.336658354 | 0.00 |
| HBB | llama-2-13b-chat | 6 | 0 | 2406 | 135 | 0.336658354 | 0.00 |
| HGF | llama-2-13b-chat | 18 | 0 | 2406 | 135 | 1.009975062 | 0.00 |
| HNF4A | llama-2-13b-chat | 6 | 0 | 2406 | 135 | 0.336658354 | 0.00 |
| HNRNPH2 | llama-2-13b-chat | 6 | 0 | 2406 | 135 | 0.336658354 | 0.00 |
| IFT172 | llama-2-13b-chat | 6 | 0 | 2406 | 135 | 0.336658354 | 0.00 |
| IQSEC2 | llama-2-13b-chat | 6 | 0 | 2406 | 135 | 0.336658354 | 0.00 |
| KAT6A | llama-2-13b-chat | 12 | 0 | 2406 | 135 | 0.673316708 | 0.00 |
| KAT6B | llama-2-13b-chat | 6 | 0 | 2406 | 135 | 0.336658354 | 0.00 |
| KCNK9 | llama-2-13b-chat | 6 | 0 | 2406 | 135 | 0.336658354 | 0.00 |
| KCNN4 | llama-2-13b-chat | 6 | 0 | 2406 | 135 | 0.336658354 | 0.00 |
| KMT2A | llama-2-13b-chat | 6 | 0 | 2406 | 135 | 0.336658354 | 0.00 |
| KMT2D | llama-2-13b-chat | 6 | 0 | 2406 | 135 | 0.336658354 | 0.00 |
| LIPT2 | llama-2-13b-chat | 36 | 0 | 2406 | 135 | 2.019950124 | 0.00 |
| MAF | llama-2-13b-chat | 6 | 0 | 2406 | 135 | 0.336658354 | 0.00 |
| MAN2B1 | llama-2-13b-chat | 6 | 0 | 2406 | 135 | 0.336658354 | 0.00 |
| MECP2 | llama-2-13b-chat | 6 | 0 | 2406 | 135 | 0.336658354 | 0.00 |
| MED13L | llama-2-13b-chat | 12 | 0 | 2406 | 135 | 0.673316708 | 0.00 |
| MEN1 | llama-2-13b-chat | 6 | 0 | 2406 | 135 | 0.336658354 | 0.00 |
| MMUT | llama-2-13b-chat | 6 | 0 | 2406 | 135 | 0.336658354 | 0.00 |
| MRPS34 | llama-2-13b-chat | 66 | 0 | 2406 | 135 | 3.703241895 | 0.00 |
| MYH10 | llama-2-13b-chat | 6 | 0 | 2406 | 135 | 0.336658354 | 0.00 |
| MYH9 | llama-2-13b-chat | 12 | 0 | 2406 | 135 | 0.673316708 | 0.00 |
| NAA15 | llama-2-13b-chat | 6 | 0 | 2406 | 135 | 0.336658354 | 0.00 |
| NPC1 | llama-2-13b-chat | 6 | 0 | 2406 | 135 | 0.336658354 | 0.00 |
| NPHS2 | llama-2-13b-chat | 6 | 0 | 2406 | 135 | 0.336658354 | 0.00 |
| NPR3 | llama-2-13b-chat | 30 | 0 | 2406 | 135 | 1.683291770 | 0.00 |
| NSD1 | llama-2-13b-chat | 6 | 0 | 2406 | 135 | 0.336658354 | 0.00 |
| OPA3 | llama-2-13b-chat | 12 | 0 | 2406 | 135 | 0.673316708 | 0.00 |
| PACS1 | llama-2-13b-chat | 12 | 0 | 2406 | 135 | 0.673316708 | 0.00 |
| PAX6 | llama-2-13b-chat | 6 | 0 | 2406 | 135 | 0.336658354 | 0.00 |
| PEPD | llama-2-13b-chat | 6 | 0 | 2406 | 135 | 0.336658354 | 0.00 |
| PEX1 | llama-2-13b-chat | 6 | 0 | 2406 | 135 | 0.336658354 | 0.00 |
| PEX26 | llama-2-13b-chat | 6 | 0 | 2406 | 135 | 0.336658354 | 0.00 |
| PHIP | llama-2-13b-chat | 18 | 0 | 2406 | 135 | 1.009975062 | 0.00 |
| PIGN | llama-2-13b-chat | 6 | 0 | 2406 | 135 | 0.336658354 | 0.00 |
| PLA2G4A | llama-2-13b-chat | 6 | 0 | 2406 | 135 | 0.336658354 | 0.00 |
| PMPCA | llama-2-13b-chat | 18 | 0 | 2406 | 135 | 1.009975062 | 0.00 |
| POMT1 | llama-2-13b-chat | 6 | 0 | 2406 | 135 | 0.336658354 | 0.00 |
| PPP3CA | llama-2-13b-chat | 12 | 0 | 2406 | 135 | 0.673316708 | 0.00 |
| PRRT2 | llama-2-13b-chat | 12 | 0 | 2406 | 135 | 0.673316708 | 0.00 |
| PUF60 | llama-2-13b-chat | 6 | 0 | 2406 | 135 | 0.336658354 | 0.00 |
| PURA | llama-2-13b-chat | 18 | 0 | 2406 | 135 | 1.009975062 | 0.00 |
| PYCR2 | llama-2-13b-chat | 6 | 0 | 2406 | 135 | 0.336658354 | 0.00 |
| RAB11B | llama-2-13b-chat | 60 | 0 | 2406 | 135 | 3.366583541 | 0.00 |
| RAC1 | llama-2-13b-chat | 66 | 0 | 2406 | 135 | 3.703241895 | 0.00 |
| REST | llama-2-13b-chat | 18 | 0 | 2406 | 135 | 1.009975062 | 0.00 |
| RET | llama-2-13b-chat | 6 | 0 | 2406 | 135 | 0.336658354 | 0.00 |
| SAMHD1 | llama-2-13b-chat | 12 | 0 | 2406 | 135 | 0.673316708 | 0.00 |
| SCNN1B | llama-2-13b-chat | 12 | 0 | 2406 | 135 | 0.673316708 | 0.00 |
| SDHA | llama-2-13b-chat | 6 | 0 | 2406 | 135 | 0.336658354 | 0.00 |
| SHH | llama-2-13b-chat | 6 | 0 | 2406 | 135 | 0.336658354 | 0.00 |
| SLC12A2 | llama-2-13b-chat | 12 | 0 | 2406 | 135 | 0.673316708 | 0.00 |
| SLC19A3 | llama-2-13b-chat | 18 | 0 | 2406 | 135 | 1.009975062 | 0.00 |
| SLC1A4 | llama-2-13b-chat | 6 | 0 | 2406 | 135 | 0.336658354 | 0.00 |
| SMARCA2 | llama-2-13b-chat | 6 | 0 | 2406 | 135 | 0.336658354 | 0.00 |
| SMC1A | llama-2-13b-chat | 12 | 0 | 2406 | 135 | 0.673316708 | 0.00 |
| SNAP25 | llama-2-13b-chat | 6 | 0 | 2406 | 135 | 0.336658354 | 0.00 |
| SON | llama-2-13b-chat | 6 | 0 | 2406 | 135 | 0.336658354 | 0.00 |
| SPART | llama-2-13b-chat | 36 | 0 | 2406 | 135 | 2.019950124 | 0.00 |
| SPTA1 | llama-2-13b-chat | 6 | 0 | 2406 | 135 | 0.336658354 | 0.00 |
| STRC | llama-2-13b-chat | 6 | 0 | 2406 | 135 | 0.336658354 | 0.00 |
| SUFU | llama-2-13b-chat | 12 | 0 | 2406 | 135 | 0.673316708 | 0.00 |
| TAF1 | llama-2-13b-chat | 84 | 0 | 2406 | 135 | 4.713216957 | 0.00 |
| TBC1D23 | llama-2-13b-chat | 18 | 0 | 2406 | 135 | 1.009975062 | 0.00 |
| TBCK | llama-2-13b-chat | 6 | 0 | 2406 | 135 | 0.336658354 | 0.00 |
| TBX3 | llama-2-13b-chat | 6 | 0 | 2406 | 135 | 0.336658354 | 0.00 |
| TBXAS1 | llama-2-13b-chat | 6 | 0 | 2406 | 135 | 0.336658354 | 0.00 |
| TCOF1 | llama-2-13b-chat | 6 | 0 | 2406 | 135 | 0.336658354 | 0.00 |
| TMEM231 | llama-2-13b-chat | 6 | 0 | 2406 | 135 | 0.336658354 | 0.00 |
| TMEM87B | llama-2-13b-chat | 12 | 0 | 2406 | 135 | 0.673316708 | 0.00 |
| TUBB3 | llama-2-13b-chat | 6 | 0 | 2406 | 135 | 0.336658354 | 0.00 |
| TWIST1 | llama-2-13b-chat | 6 | 0 | 2406 | 135 | 0.336658354 | 0.00 |
| TWNK | llama-2-13b-chat | 12 | 0 | 2406 | 135 | 0.673316708 | 0.00 |
| UNC80 | llama-2-13b-chat | 6 | 0 | 2406 | 135 | 0.336658354 | 0.00 |
| VARS1 | llama-2-13b-chat | 18 | 0 | 2406 | 135 | 1.009975062 | 0.00 |
| WDR26 | llama-2-13b-chat | 156 | 0 | 2406 | 135 | 8.753117206 | 0.00 |
| WFS1 | llama-2-13b-chat | 6 | 0 | 2406 | 135 | 0.336658354 | 0.00 |
| WWOX | llama-2-13b-chat | 6 | 0 | 2406 | 135 | 0.336658354 | 0.00 |
| YWHAG | llama-2-13b-chat | 72 | 0 | 2406 | 135 | 4.039900249 | 0.00 |
| ZMYND11 | llama-2-13b-chat | 12 | 0 | 2406 | 135 | 0.673316708 | 0.00 |
| BRCA1 | gpt-4 | 24 | 24 | 9624 | 1448 | 3.610972568 | 6.646408839 |
| CYP17A1 | gpt-4 | 24 | 24 | 9624 | 1448 | 3.610972568 | 6.646408839 |
| F8 | gpt-4 | 24 | 24 | 9624 | 1448 | 3.610972568 | 6.646408839 |
| IL12RB1 | gpt-4 | 24 | 24 | 9624 | 1448 | 3.610972568 | 6.646408839 |
| PAX6 | gpt-4 | 24 | 24 | 9624 | 1448 | 3.610972568 | 6.646408839 |
| PEPD | gpt-4 | 24 | 24 | 9624 | 1448 | 3.610972568 | 6.646408839 |
| SHH | gpt-4 | 24 | 24 | 9624 | 1448 | 3.610972568 | 6.646408839 |
| SPG11 | gpt-4 | 24 | 24 | 9624 | 1448 | 3.610972568 | 6.646408839 |
| TP53 | gpt-4 | 336 | 335 | 9624 | 1448 | 50.55361596 | 6.626627861 |
| ANK1 | gpt-4 | 24 | 23 | 9624 | 1448 | 3.610972568 | 6.369475138 |
| TSHR | gpt-4 | 24 | 23 | 9624 | 1448 | 3.610972568 | 6.369475138 |
| SCN1A | gpt-4 | 48 | 45 | 9624 | 1448 | 7.221945137 | 6.231008287 |
| ACTG2 | gpt-4 | 24 | 22 | 9624 | 1448 | 3.610972568 | 6.092541436 |
| COL10A1 | gpt-4 | 24 | 22 | 9624 | 1448 | 3.610972568 | 6.092541436 |
| MMUT | gpt-4 | 24 | 22 | 9624 | 1448 | 3.610972568 | 6.092541436 |
| CDKL5 | gpt-4 | 96 | 87 | 9624 | 1448 | 14.44389027 | 6.023308011 |
| BTK | gpt-4 | 24 | 21 | 9624 | 1448 | 3.610972568 | 5.82 |
| SPTA1 | gpt-4 | 24 | 21 | 9624 | 1448 | 3.610972568 | 5.82 |
| AIRE | gpt-4 | 48 | 40 | 9624 | 1448 | 7.221945137 | 5.538674033 |
| MECP2 | gpt-4 | 24 | 20 | 9624 | 1448 | 3.610972568 | 5.538674033 |
| BRAF | gpt-4 | 96 | 77 | 9624 | 1448 | 14.44389027 | 5.330973756 |
| LMNA | gpt-4 | 24 | 19 | 9624 | 1448 | 3.610972568 | 5.261740331 |
| MYO7A | gpt-4 | 24 | 19 | 9624 | 1448 | 3.610972568 | 5.261740331 |
| WFS1 | gpt-4 | 24 | 19 | 9624 | 1448 | 3.610972568 | 5.261740331 |
| SOX9 | gpt-4 | 48 | 35 | 9624 | 1448 | 7.221945137 | 4.846339779 |
| GABRA1 | gpt-4 | 24 | 17 | 9624 | 1448 | 3.610972568 | 4.707872928 |
| RYR1 | gpt-4 | 24 | 17 | 9624 | 1448 | 3.610972568 | 4.707872928 |
| SCN8A | gpt-4 | 48 | 28 | 9624 | 1448 | 7.221945137 | 3.88 |
| SCNN1B | gpt-4 | 48 | 27 | 9624 | 1448 | 7.221945137 | 3.738604972 |
| APTX | gpt-4 | 24 | 12 | 9624 | 1448 | 3.610972568 | 3.323204419 |
| ASXL1 | gpt-4 | 24 | 12 | 9624 | 1448 | 3.610972568 | 3.323204419 |
| PTEN | gpt-4 | 48 | 24 | 9624 | 1448 | 7.221945137 | 3.323204419 |
| RET | gpt-4 | 24 | 11 | 9624 | 1448 | 3.610972568 | 3.046270718 |
| TWIST1 | gpt-4 | 24 | 11 | 9624 | 1448 | 3.610972568 | 3.046270718 |
| ABCD1 | gpt-4 | 24 | 10 | 9624 | 1448 | 3.610972568 | 2.769337016 |
| GRIN2B | gpt-4 | 24 | 10 | 9624 | 1448 | 3.610972568 | 2.769337016 |
| SYNGAP1 | gpt-4 | 72 | 26 | 9624 | 1448 | 10.83291770 | 2.400092081 |
| TWNK | gpt-4 | 48 | 16 | 9624 | 1448 | 7.221945137 | 2.215469613 |
| MED13L | gpt-4 | 48 | 12 | 9624 | 1448 | 7.221945137 | 1.661602209 |
| SDHA | gpt-4 | 24 | 6 | 9624 | 1448 | 3.610972568 | 1.661602209 |
| ARID1B | gpt-4 | 96 | 21 | 9624 | 1448 | 14.44389027 | 1.453901933 |
| KMT2A | gpt-4 | 24 | 5 | 9624 | 1448 | 3.610972568 | 1.384668508 |
| WT1 | gpt-4 | 24 | 5 | 9624 | 1448 | 3.610972568 | 1.384668508 |
| SLC19A3 | gpt-4 | 72 | 14 | 9624 | 1448 | 10.83291770 | 1.292357274 |
| OPA3 | gpt-4 | 48 | 9 | 9624 | 1448 | 7.221945137 | 1.246201657 |
| PTPN11 | gpt-4 | 48 | 9 | 9624 | 1448 | 7.221945137 | 1.246201657 |
| ARX | gpt-4 | 24 | 4 | 9624 | 1448 | 3.610972568 | 1.107734806 |
| DDX3X | gpt-4 | 24 | 4 | 9624 | 1448 | 3.610972568 | 1.107734806 |
| IQSEC2 | gpt-4 | 24 | 4 | 9624 | 1448 | 3.610972568 | 1.107734806 |
| TCOF1 | gpt-4 | 24 | 4 | 9624 | 1448 | 3.610972568 | 1.107734806 |
| ANKRD11 | gpt-4 | 96 | 12 | 9624 | 1448 | 14.44389027 | 0.830801104 |
| CCN6 | gpt-4 | 24 | 3 | 9624 | 1448 | 3.610972568 | 0.830801104 |
| KCNN4 | gpt-4 | 24 | 3 | 9624 | 1448 | 3.610972568 | 0.830801104 |
| MEN1 | gpt-4 | 24 | 3 | 9624 | 1448 | 3.610972568 | 0.830801104 |
| PRRT2 | gpt-4 | 48 | 6 | 9624 | 1448 | 7.221945137 | 0.830801104 |
| EFL1 | gpt-4 | 48 | 5 | 9624 | 1448 | 7.221945137 | 0.692334254 |
| ASXL3 | gpt-4 | 72 | 6 | 9624 | 1448 | 10.83291770 | 0.553867403 |
| COL18A1 | gpt-4 | 24 | 2 | 9624 | 1448 | 3.610972568 | 0.553867403 |
| CTSA | gpt-4 | 24 | 2 | 9624 | 1448 | 3.610972568 | 0.553867403 |
| IFT172 | gpt-4 | 24 | 2 | 9624 | 1448 | 3.610972568 | 0.553867403 |
| NPC1 | gpt-4 | 24 | 2 | 9624 | 1448 | 3.610972568 | 0.553867403 |
| SAMHD1 | gpt-4 | 48 | 4 | 9624 | 1448 | 7.221945137 | 0.553867403 |
| SPART | gpt-4 | 144 | 11 | 9624 | 1448 | 21.66583541 | 0.507711786 |
| EHMT1 | gpt-4 | 48 | 3 | 9624 | 1448 | 7.221945137 | 0.415400552 |
| PACS1 | gpt-4 | 48 | 3 | 9624 | 1448 | 7.221945137 | 0.415400552 |
| LIPT2 | gpt-4 | 144 | 7 | 9624 | 1448 | 21.66583541 | 0.323089318 |
| ATN1 | gpt-4 | 24 | 1 | 9624 | 1448 | 3.610972568 | 0.276933701 |
| ATRX | gpt-4 | 24 | 1 | 9624 | 1448 | 3.610972568 | 0.276933701 |
| PEX1 | gpt-4 | 24 | 1 | 9624 | 1448 | 3.610972568 | 0.276933701 |
| SMC1A | gpt-4 | 48 | 2 | 9624 | 1448 | 7.221945137 | 0.276933701 |
| SUFU | gpt-4 | 48 | 2 | 9624 | 1448 | 7.221945137 | 0.276933701 |
| TMEM231 | gpt-4 | 24 | 1 | 9624 | 1448 | 3.610972568 | 0.276933701 |
| ATP1A3 | gpt-4 | 72 | 2 | 9624 | 1448 | 10.83291770 | 0.184622467 |
| AIFM1 | gpt-4 | 48 | 1 | 9624 | 1448 | 7.221945137 | 0.138466850 |
| ALPK3 | gpt-4 | 48 | 1 | 9624 | 1448 | 7.221945137 | 0.138466850 |
| CLUAP1 | gpt-4 | 48 | 1 | 9624 | 1448 | 7.221945137 | 0.138466850 |
| PURA | gpt-4 | 72 | 1 | 9624 | 1448 | 10.83291770 | 0.09 |
| ADA2 | gpt-4 | 24 | 0 | 9624 | 1448 | 3.610972568 | 0.00 |
| ADSS1 | gpt-4 | 24 | 0 | 9624 | 1448 | 3.610972568 | 0.00 |
| AGL | gpt-4 | 24 | 0 | 9624 | 1448 | 3.610972568 | 0.00 |
| AHDC1 | gpt-4 | 72 | 0 | 9624 | 1448 | 10.83291770 | 0.00 |
| ALG13 | gpt-4 | 24 | 0 | 9624 | 1448 | 3.610972568 | 0.00 |
| AQP4 | gpt-4 | 48 | 0 | 9624 | 1448 | 7.221945137 | 0.00 |
| ARID2 | gpt-4 | 24 | 0 | 9624 | 1448 | 3.610972568 | 0.00 |
| AUTS2 | gpt-4 | 24 | 0 | 9624 | 1448 | 3.610972568 | 0.00 |
| B3GLCT | gpt-4 | 24 | 0 | 9624 | 1448 | 3.610972568 | 0.00 |
| BICD2 | gpt-4 | 96 | 0 | 9624 | 1448 | 14.44389027 | 0.00 |
| BTD | gpt-4 | 24 | 0 | 9624 | 1448 | 3.610972568 | 0.00 |
| C3AR1 | gpt-4 | 48 | 0 | 9624 | 1448 | 7.221945137 | 0.00 |
| CACNA1C | gpt-4 | 24 | 0 | 9624 | 1448 | 3.610972568 | 0.00 |
| CAMK4 | gpt-4 | 48 | 0 | 9624 | 1448 | 7.221945137 | 0.00 |
| CAPN5 | gpt-4 | 48 | 0 | 9624 | 1448 | 7.221945137 | 0.00 |
| CDK10 | gpt-4 | 384 | 0 | 9624 | 1448 | 57.77556109 | 0.00 |
| CDK13 | gpt-4 | 24 | 0 | 9624 | 1448 | 3.610972568 | 0.00 |
| CHAMP1 | gpt-4 | 24 | 0 | 9624 | 1448 | 3.610972568 | 0.00 |
| COL4A1 | gpt-4 | 24 | 0 | 9624 | 1448 | 3.610972568 | 0.00 |
| COL7A1 | gpt-4 | 48 | 0 | 9624 | 1448 | 7.221945137 | 0.00 |
| CTNNB1 | gpt-4 | 24 | 0 | 9624 | 1448 | 3.610972568 | 0.00 |
| DCAF17 | gpt-4 | 24 | 0 | 9624 | 1448 | 3.610972568 | 0.00 |
| DDR2 | gpt-4 | 24 | 0 | 9624 | 1448 | 3.610972568 | 0.00 |
| DDX11 | gpt-4 | 24 | 0 | 9624 | 1448 | 3.610972568 | 0.00 |
| DHX30 | gpt-4 | 576 | 0 | 9624 | 1448 | 86.66334164 | 0.00 |
| EBF3 | gpt-4 | 24 | 0 | 9624 | 1448 | 3.610972568 | 0.00 |
| EGFR | gpt-4 | 48 | 0 | 9624 | 1448 | 7.221945137 | 0.00 |
| EIF2AK4 | gpt-4 | 24 | 0 | 9624 | 1448 | 3.610972568 | 0.00 |
| EP300 | gpt-4 | 24 | 0 | 9624 | 1448 | 3.610972568 | 0.00 |
| F13A1 | gpt-4 | 24 | 0 | 9624 | 1448 | 3.610972568 | 0.00 |
| FDXR | gpt-4 | 336 | 0 | 9624 | 1448 | 50.55361596 | 0.00 |
| FGD1 | gpt-4 | 48 | 0 | 9624 | 1448 | 7.221945137 | 0.00 |
| FOXP2 | gpt-4 | 48 | 0 | 9624 | 1448 | 7.221945137 | 0.00 |
| G6PD | gpt-4 | 24 | 0 | 9624 | 1448 | 3.610972568 | 0.00 |
| GNAS | gpt-4 | 24 | 0 | 9624 | 1448 | 3.610972568 | 0.00 |
| GRM1 | gpt-4 | 96 | 0 | 9624 | 1448 | 14.44389027 | 0.00 |
| HBB | gpt-4 | 24 | 0 | 9624 | 1448 | 3.610972568 | 0.00 |
| HGF | gpt-4 | 72 | 0 | 9624 | 1448 | 10.83291770 | 0.00 |
| HNF4A | gpt-4 | 24 | 0 | 9624 | 1448 | 3.610972568 | 0.00 |
| HNRNPH2 | gpt-4 | 24 | 0 | 9624 | 1448 | 3.610972568 | 0.00 |
| KAT6A | gpt-4 | 48 | 0 | 9624 | 1448 | 7.221945137 | 0.00 |
| KAT6B | gpt-4 | 24 | 0 | 9624 | 1448 | 3.610972568 | 0.00 |
| KCNK9 | gpt-4 | 24 | 0 | 9624 | 1448 | 3.610972568 | 0.00 |
| KMT2D | gpt-4 | 24 | 0 | 9624 | 1448 | 3.610972568 | 0.00 |
| MAF | gpt-4 | 24 | 0 | 9624 | 1448 | 3.610972568 | 0.00 |
| MAN2B1 | gpt-4 | 24 | 0 | 9624 | 1448 | 3.610972568 | 0.00 |
| MRPS34 | gpt-4 | 264 | 0 | 9624 | 1448 | 39.72069825 | 0.00 |
| MYH10 | gpt-4 | 24 | 0 | 9624 | 1448 | 3.610972568 | 0.00 |
| MYH9 | gpt-4 | 48 | 0 | 9624 | 1448 | 7.221945137 | 0.00 |
| NAA15 | gpt-4 | 24 | 0 | 9624 | 1448 | 3.610972568 | 0.00 |
| NPHS2 | gpt-4 | 24 | 0 | 9624 | 1448 | 3.610972568 | 0.00 |
| NPR3 | gpt-4 | 120 | 0 | 9624 | 1448 | 18.05486284 | 0.00 |
| NR2F1 | gpt-4 | 96 | 0 | 9624 | 1448 | 14.44389027 | 0.00 |
| NSD1 | gpt-4 | 24 | 0 | 9624 | 1448 | 3.610972568 | 0.00 |
| PEX26 | gpt-4 | 24 | 0 | 9624 | 1448 | 3.610972568 | 0.00 |
| PHIP | gpt-4 | 72 | 0 | 9624 | 1448 | 10.83291770 | 0.00 |
| PIGN | gpt-4 | 24 | 0 | 9624 | 1448 | 3.610972568 | 0.00 |
| PLA2G4A | gpt-4 | 24 | 0 | 9624 | 1448 | 3.610972568 | 0.00 |
| PMM2 | gpt-4 | 48 | 0 | 9624 | 1448 | 7.221945137 | 0.00 |
| PMPCA | gpt-4 | 72 | 0 | 9624 | 1448 | 10.83291770 | 0.00 |
| POMT1 | gpt-4 | 24 | 0 | 9624 | 1448 | 3.610972568 | 0.00 |
| PPP3CA | gpt-4 | 48 | 0 | 9624 | 1448 | 7.221945137 | 0.00 |
| PUF60 | gpt-4 | 24 | 0 | 9624 | 1448 | 3.610972568 | 0.00 |
| PYCR2 | gpt-4 | 24 | 0 | 9624 | 1448 | 3.610972568 | 0.00 |
| RAB11B | gpt-4 | 240 | 0 | 9624 | 1448 | 36.10972568 | 0.00 |
| RAC1 | gpt-4 | 264 | 0 | 9624 | 1448 | 39.72069825 | 0.00 |
| REST | gpt-4 | 72 | 0 | 9624 | 1448 | 10.83291770 | 0.00 |
| SLC12A2 | gpt-4 | 48 | 0 | 9624 | 1448 | 7.221945137 | 0.00 |
| SLC1A4 | gpt-4 | 24 | 0 | 9624 | 1448 | 3.610972568 | 0.00 |
| SMARCA2 | gpt-4 | 24 | 0 | 9624 | 1448 | 3.610972568 | 0.00 |
| SMARCA4 | gpt-4 | 24 | 0 | 9624 | 1448 | 3.610972568 | 0.00 |
| SNAP25 | gpt-4 | 24 | 0 | 9624 | 1448 | 3.610972568 | 0.00 |
| SON | gpt-4 | 24 | 0 | 9624 | 1448 | 3.610972568 | 0.00 |
| STRC | gpt-4 | 24 | 0 | 9624 | 1448 | 3.610972568 | 0.00 |
| TAF1 | gpt-4 | 336 | 0 | 9624 | 1448 | 50.55361596 | 0.00 |
| TBC1D23 | gpt-4 | 72 | 0 | 9624 | 1448 | 10.83291770 | 0.00 |
| TBCK | gpt-4 | 24 | 0 | 9624 | 1448 | 3.610972568 | 0.00 |
| TBX3 | gpt-4 | 24 | 0 | 9624 | 1448 | 3.610972568 | 0.00 |
| TBXAS1 | gpt-4 | 24 | 0 | 9624 | 1448 | 3.610972568 | 0.00 |
| TMEM87B | gpt-4 | 48 | 0 | 9624 | 1448 | 7.221945137 | 0.00 |
| TUBB3 | gpt-4 | 24 | 0 | 9624 | 1448 | 3.610972568 | 0.00 |
| UNC80 | gpt-4 | 24 | 0 | 9624 | 1448 | 3.610972568 | 0.00 |
| VARS1 | gpt-4 | 72 | 0 | 9624 | 1448 | 10.83291770 | 0.00 |
| WDR45 | gpt-4 | 24 | 0 | 9624 | 1448 | 3.610972568 | 0.00 |
| WWOX | gpt-4 | 24 | 0 | 9624 | 1448 | 3.610972568 | 0.00 |
| YWHAG | gpt-4 | 288 | 0 | 9624 | 1448 | 43.33167082 | 0.00 |
| ZMYND11 | gpt-4 | 48 | 0 | 9624 | 1448 | 7.221945137 | 0.00 |
| ANK1 | gpt-3.5-turbo | 24 | 20 | 9624 | 786 | 1.960099750 | 10.20356234 |
| BRCA1 | gpt-3.5-turbo | 24 | 18 | 9624 | 786 | 1.960099750 | 9.183206106 |
| PEPD | gpt-3.5-turbo | 24 | 18 | 9624 | 786 | 1.960099750 | 9.183206106 |
| BTK | gpt-3.5-turbo | 24 | 16 | 9624 | 786 | 1.960099750 | 8.162849872 |
| SOX9 | gpt-3.5-turbo | 48 | 32 | 9624 | 786 | 3.920199501 | 8.162849872 |
| SCN1A | gpt-3.5-turbo | 48 | 31 | 9624 | 786 | 3.920199501 | 7.907760814 |
| TP53 | gpt-3.5-turbo | 336 | 215 | 9624 | 786 | 27.44139650 | 7.83 |
| NSD1 | gpt-3.5-turbo | 24 | 14 | 9624 | 786 | 1.960099750 | 7.142493638 |
| SPTA1 | gpt-3.5-turbo | 24 | 13 | 9624 | 786 | 1.960099750 | 6.632315521 |
| CDKL5 | gpt-3.5-turbo | 96 | 48 | 9624 | 786 | 7.840399002 | 6.122137404 |
| MECP2 | gpt-3.5-turbo | 24 | 12 | 9624 | 786 | 1.960099750 | 6.122137404 |
| TSHR | gpt-3.5-turbo | 24 | 12 | 9624 | 786 | 1.960099750 | 6.122137404 |
| MMUT | gpt-3.5-turbo | 24 | 11 | 9624 | 786 | 1.960099750 | 5.611959287 |
| SCNN1B | gpt-3.5-turbo | 48 | 21 | 9624 | 786 | 3.920199501 | 5.356870229 |
| ACTG2 | gpt-3.5-turbo | 24 | 10 | 9624 | 786 | 1.960099750 | 5.10 |
| BRAF | gpt-3.5-turbo | 96 | 36 | 9624 | 786 | 7.840399002 | 4.591603053 |
| DDX3X | gpt-3.5-turbo | 24 | 9 | 9624 | 786 | 1.960099750 | 4.591603053 |
| LMNA | gpt-3.5-turbo | 24 | 9 | 9624 | 786 | 1.960099750 | 4.591603053 |
| RYR1 | gpt-3.5-turbo | 24 | 9 | 9624 | 786 | 1.960099750 | 4.591603053 |
| WT1 | gpt-3.5-turbo | 24 | 9 | 9624 | 786 | 1.960099750 | 4.591603053 |
| SHH | gpt-3.5-turbo | 24 | 8 | 9624 | 786 | 1.960099750 | 4.081424936 |
| MYO7A | gpt-3.5-turbo | 24 | 7 | 9624 | 786 | 1.960099750 | 3.571246819 |
| PAX6 | gpt-3.5-turbo | 24 | 7 | 9624 | 786 | 1.960099750 | 3.571246819 |
| ABCD1 | gpt-3.5-turbo | 24 | 6 | 9624 | 786 | 1.960099750 | 3.061068702 |
| APTX | gpt-3.5-turbo | 24 | 6 | 9624 | 786 | 1.960099750 | 3.061068702 |
| ASXL1 | gpt-3.5-turbo | 24 | 6 | 9624 | 786 | 1.960099750 | 3.061068702 |
| COL4A1 | gpt-3.5-turbo | 24 | 6 | 9624 | 786 | 1.960099750 | 3.061068702 |
| MEN1 | gpt-3.5-turbo | 24 | 6 | 9624 | 786 | 1.960099750 | 3.061068702 |
| PTEN | gpt-3.5-turbo | 48 | 12 | 9624 | 786 | 3.920199501 | 3.061068702 |
| SDHA | gpt-3.5-turbo | 24 | 6 | 9624 | 786 | 1.960099750 | 3.061068702 |
| TWIST1 | gpt-3.5-turbo | 24 | 6 | 9624 | 786 | 1.960099750 | 3.061068702 |
| ARID1B | gpt-3.5-turbo | 96 | 23 | 9624 | 786 | 7.840399002 | 2.93 |
| GABRA1 | gpt-3.5-turbo | 24 | 5 | 9624 | 786 | 1.960099750 | 2.55 |
| SPG11 | gpt-3.5-turbo | 24 | 5 | 9624 | 786 | 1.960099750 | 2.55 |
| SYNGAP1 | gpt-3.5-turbo | 72 | 14 | 9624 | 786 | 5.880299251 | 2.380831212 |
| EHMT1 | gpt-3.5-turbo | 48 | 9 | 9624 | 786 | 3.920199501 | 2.295801526 |
| CCN6 | gpt-3.5-turbo | 24 | 4 | 9624 | 786 | 1.960099750 | 2.040712468 |
| MED13L | gpt-3.5-turbo | 48 | 8 | 9624 | 786 | 3.920199501 | 2.040712468 |
| NPC1 | gpt-3.5-turbo | 24 | 4 | 9624 | 786 | 1.960099750 | 2.040712468 |
| SCN8A | gpt-3.5-turbo | 48 | 7 | 9624 | 786 | 3.920199501 | 1.785623409 |
| ASXL3 | gpt-3.5-turbo | 72 | 9 | 9624 | 786 | 5.880299251 | 1.530534351 |
| COL10A1 | gpt-3.5-turbo | 24 | 3 | 9624 | 786 | 1.960099750 | 1.530534351 |
| F8 | gpt-3.5-turbo | 24 | 3 | 9624 | 786 | 1.960099750 | 1.530534351 |
| HNRNPH2 | gpt-3.5-turbo | 24 | 3 | 9624 | 786 | 1.960099750 | 1.530534351 |
| IL12RB1 | gpt-3.5-turbo | 24 | 3 | 9624 | 786 | 1.960099750 | 1.530534351 |
| GRM1 | gpt-3.5-turbo | 96 | 11 | 9624 | 786 | 7.840399002 | 1.402989821 |
| KAT6A | gpt-3.5-turbo | 48 | 5 | 9624 | 786 | 3.920199501 | 1.275445292 |
| ADA2 | gpt-3.5-turbo | 24 | 2 | 9624 | 786 | 1.960099750 | 1.020356234 |
| SMARCA4 | gpt-3.5-turbo | 24 | 2 | 9624 | 786 | 1.960099750 | 1.020356234 |
| BICD2 | gpt-3.5-turbo | 96 | 7 | 9624 | 786 | 7.840399002 | 0.892811704 |
| MYH9 | gpt-3.5-turbo | 48 | 3 | 9624 | 786 | 3.920199501 | 0.765267175 |
| PRRT2 | gpt-3.5-turbo | 48 | 3 | 9624 | 786 | 3.920199501 | 0.765267175 |
| TWNK | gpt-3.5-turbo | 48 | 3 | 9624 | 786 | 3.920199501 | 0.765267175 |
| ANKRD11 | gpt-3.5-turbo | 96 | 5 | 9624 | 786 | 7.840399002 | 0.637722646 |
| AIRE | gpt-3.5-turbo | 48 | 2 | 9624 | 786 | 3.920199501 | 0.51 |
| CACNA1C | gpt-3.5-turbo | 24 | 1 | 9624 | 786 | 1.960099750 | 0.51 |
| PTPN11 | gpt-3.5-turbo | 48 | 2 | 9624 | 786 | 3.920199501 | 0.51 |
| EGFR | gpt-3.5-turbo | 48 | 1 | 9624 | 786 | 3.920199501 | 0.26 |
| ADSS1 | gpt-3.5-turbo | 24 | 0 | 9624 | 786 | 1.960099750 | 0.00 |
| AGL | gpt-3.5-turbo | 24 | 0 | 9624 | 786 | 1.960099750 | 0.00 |
| AHDC1 | gpt-3.5-turbo | 72 | 0 | 9624 | 786 | 5.880299251 | 0.00 |
| AIFM1 | gpt-3.5-turbo | 48 | 0 | 9624 | 786 | 3.920199501 | 0.00 |
| ALG13 | gpt-3.5-turbo | 24 | 0 | 9624 | 786 | 1.960099750 | 0.00 |
| ALPK3 | gpt-3.5-turbo | 48 | 0 | 9624 | 786 | 3.920199501 | 0.00 |
| AQP4 | gpt-3.5-turbo | 48 | 0 | 9624 | 786 | 3.920199501 | 0.00 |
| ARID2 | gpt-3.5-turbo | 24 | 0 | 9624 | 786 | 1.960099750 | 0.00 |
| ARX | gpt-3.5-turbo | 24 | 0 | 9624 | 786 | 1.960099750 | 0.00 |
| ATN1 | gpt-3.5-turbo | 24 | 0 | 9624 | 786 | 1.960099750 | 0.00 |
| ATP1A3 | gpt-3.5-turbo | 72 | 0 | 9624 | 786 | 5.880299251 | 0.00 |
| ATRX | gpt-3.5-turbo | 24 | 0 | 9624 | 786 | 1.960099750 | 0.00 |
| AUTS2 | gpt-3.5-turbo | 24 | 0 | 9624 | 786 | 1.960099750 | 0.00 |
| B3GLCT | gpt-3.5-turbo | 24 | 0 | 9624 | 786 | 1.960099750 | 0.00 |
| BTD | gpt-3.5-turbo | 24 | 0 | 9624 | 786 | 1.960099750 | 0.00 |
| C3AR1 | gpt-3.5-turbo | 48 | 0 | 9624 | 786 | 3.920199501 | 0.00 |
| CAMK4 | gpt-3.5-turbo | 48 | 0 | 9624 | 786 | 3.920199501 | 0.00 |
| CAPN5 | gpt-3.5-turbo | 48 | 0 | 9624 | 786 | 3.920199501 | 0.00 |
| CDK10 | gpt-3.5-turbo | 384 | 0 | 9624 | 786 | 31.36159600 | 0.00 |
| CDK13 | gpt-3.5-turbo | 24 | 0 | 9624 | 786 | 1.960099750 | 0.00 |
| CHAMP1 | gpt-3.5-turbo | 24 | 0 | 9624 | 786 | 1.960099750 | 0.00 |
| CLUAP1 | gpt-3.5-turbo | 48 | 0 | 9624 | 786 | 3.920199501 | 0.00 |
| COL18A1 | gpt-3.5-turbo | 24 | 0 | 9624 | 786 | 1.960099750 | 0.00 |
| COL7A1 | gpt-3.5-turbo | 48 | 0 | 9624 | 786 | 3.920199501 | 0.00 |
| CTNNB1 | gpt-3.5-turbo | 24 | 0 | 9624 | 786 | 1.960099750 | 0.00 |
| CTSA | gpt-3.5-turbo | 24 | 0 | 9624 | 786 | 1.960099750 | 0.00 |
| CYP17A1 | gpt-3.5-turbo | 24 | 0 | 9624 | 786 | 1.960099750 | 0.00 |
| DCAF17 | gpt-3.5-turbo | 24 | 0 | 9624 | 786 | 1.960099750 | 0.00 |
| DDR2 | gpt-3.5-turbo | 24 | 0 | 9624 | 786 | 1.960099750 | 0.00 |
| DDX11 | gpt-3.5-turbo | 24 | 0 | 9624 | 786 | 1.960099750 | 0.00 |
| DHX30 | gpt-3.5-turbo | 576 | 0 | 9624 | 786 | 47.04239401 | 0.00 |
| EBF3 | gpt-3.5-turbo | 24 | 0 | 9624 | 786 | 1.960099750 | 0.00 |
| EFL1 | gpt-3.5-turbo | 48 | 0 | 9624 | 786 | 3.920199501 | 0.00 |
| EIF2AK4 | gpt-3.5-turbo | 24 | 0 | 9624 | 786 | 1.960099750 | 0.00 |
| EP300 | gpt-3.5-turbo | 24 | 0 | 9624 | 786 | 1.960099750 | 0.00 |
| F13A1 | gpt-3.5-turbo | 24 | 0 | 9624 | 786 | 1.960099750 | 0.00 |
| FDXR | gpt-3.5-turbo | 336 | 0 | 9624 | 786 | 27.44139650 | 0.00 |
| FGD1 | gpt-3.5-turbo | 48 | 0 | 9624 | 786 | 3.920199501 | 0.00 |
| FOXP2 | gpt-3.5-turbo | 48 | 0 | 9624 | 786 | 3.920199501 | 0.00 |
| G6PD | gpt-3.5-turbo | 24 | 0 | 9624 | 786 | 1.960099750 | 0.00 |
| GNAS | gpt-3.5-turbo | 24 | 0 | 9624 | 786 | 1.960099750 | 0.00 |
| GRIN2B | gpt-3.5-turbo | 24 | 0 | 9624 | 786 | 1.960099750 | 0.00 |
| HBB | gpt-3.5-turbo | 24 | 0 | 9624 | 786 | 1.960099750 | 0.00 |
| HGF | gpt-3.5-turbo | 72 | 0 | 9624 | 786 | 5.880299251 | 0.00 |
| HNF4A | gpt-3.5-turbo | 24 | 0 | 9624 | 786 | 1.960099750 | 0.00 |
| IFT172 | gpt-3.5-turbo | 24 | 0 | 9624 | 786 | 1.960099750 | 0.00 |
| IQSEC2 | gpt-3.5-turbo | 24 | 0 | 9624 | 786 | 1.960099750 | 0.00 |
| KAT6B | gpt-3.5-turbo | 24 | 0 | 9624 | 786 | 1.960099750 | 0.00 |
| KCNK9 | gpt-3.5-turbo | 24 | 0 | 9624 | 786 | 1.960099750 | 0.00 |
| KCNN4 | gpt-3.5-turbo | 24 | 0 | 9624 | 786 | 1.960099750 | 0.00 |
| KMT2A | gpt-3.5-turbo | 24 | 0 | 9624 | 786 | 1.960099750 | 0.00 |
| KMT2D | gpt-3.5-turbo | 24 | 0 | 9624 | 786 | 1.960099750 | 0.00 |
| LIPT2 | gpt-3.5-turbo | 144 | 0 | 9624 | 786 | 11.76059850 | 0.00 |
| MAF | gpt-3.5-turbo | 24 | 0 | 9624 | 786 | 1.960099750 | 0.00 |
| MAN2B1 | gpt-3.5-turbo | 24 | 0 | 9624 | 786 | 1.960099750 | 0.00 |
| MRPS34 | gpt-3.5-turbo | 264 | 0 | 9624 | 786 | 21.56109725 | 0.00 |
| MYH10 | gpt-3.5-turbo | 24 | 0 | 9624 | 786 | 1.960099750 | 0.00 |
| NAA15 | gpt-3.5-turbo | 24 | 0 | 9624 | 786 | 1.960099750 | 0.00 |
| NPHS2 | gpt-3.5-turbo | 24 | 0 | 9624 | 786 | 1.960099750 | 0.00 |
| NPR3 | gpt-3.5-turbo | 120 | 0 | 9624 | 786 | 9.800498753 | 0.00 |
| NR2F1 | gpt-3.5-turbo | 96 | 0 | 9624 | 786 | 7.840399002 | 0.00 |
| OPA3 | gpt-3.5-turbo | 48 | 0 | 9624 | 786 | 3.920199501 | 0.00 |
| PACS1 | gpt-3.5-turbo | 48 | 0 | 9624 | 786 | 3.920199501 | 0.00 |
| PEX1 | gpt-3.5-turbo | 24 | 0 | 9624 | 786 | 1.960099750 | 0.00 |
| PEX26 | gpt-3.5-turbo | 24 | 0 | 9624 | 786 | 1.960099750 | 0.00 |
| PHIP | gpt-3.5-turbo | 72 | 0 | 9624 | 786 | 5.880299251 | 0.00 |
| PIGN | gpt-3.5-turbo | 24 | 0 | 9624 | 786 | 1.960099750 | 0.00 |
| PLA2G4A | gpt-3.5-turbo | 24 | 0 | 9624 | 786 | 1.960099750 | 0.00 |
| PMM2 | gpt-3.5-turbo | 48 | 0 | 9624 | 786 | 3.920199501 | 0.00 |
| PMPCA | gpt-3.5-turbo | 72 | 0 | 9624 | 786 | 5.880299251 | 0.00 |
| POMT1 | gpt-3.5-turbo | 24 | 0 | 9624 | 786 | 1.960099750 | 0.00 |
| PPP3CA | gpt-3.5-turbo | 48 | 0 | 9624 | 786 | 3.920199501 | 0.00 |
| PUF60 | gpt-3.5-turbo | 24 | 0 | 9624 | 786 | 1.960099750 | 0.00 |
| PURA | gpt-3.5-turbo | 72 | 0 | 9624 | 786 | 5.880299251 | 0.00 |
| PYCR2 | gpt-3.5-turbo | 24 | 0 | 9624 | 786 | 1.960099750 | 0.00 |
| RAB11B | gpt-3.5-turbo | 240 | 0 | 9624 | 786 | 19.60099750 | 0.00 |
| RAC1 | gpt-3.5-turbo | 264 | 0 | 9624 | 786 | 21.56109725 | 0.00 |
| REST | gpt-3.5-turbo | 72 | 0 | 9624 | 786 | 5.880299251 | 0.00 |
| RET | gpt-3.5-turbo | 24 | 0 | 9624 | 786 | 1.960099750 | 0.00 |
| SAMHD1 | gpt-3.5-turbo | 48 | 0 | 9624 | 786 | 3.920199501 | 0.00 |
| SLC12A2 | gpt-3.5-turbo | 48 | 0 | 9624 | 786 | 3.920199501 | 0.00 |
| SLC19A3 | gpt-3.5-turbo | 72 | 0 | 9624 | 786 | 5.880299251 | 0.00 |
| SLC1A4 | gpt-3.5-turbo | 24 | 0 | 9624 | 786 | 1.960099750 | 0.00 |
| SMARCA2 | gpt-3.5-turbo | 24 | 0 | 9624 | 786 | 1.960099750 | 0.00 |
| SMC1A | gpt-3.5-turbo | 48 | 0 | 9624 | 786 | 3.920199501 | 0.00 |
| SNAP25 | gpt-3.5-turbo | 24 | 0 | 9624 | 786 | 1.960099750 | 0.00 |
| SON | gpt-3.5-turbo | 24 | 0 | 9624 | 786 | 1.960099750 | 0.00 |
| SPART | gpt-3.5-turbo | 144 | 0 | 9624 | 786 | 11.76059850 | 0.00 |
| STRC | gpt-3.5-turbo | 24 | 0 | 9624 | 786 | 1.960099750 | 0.00 |
| SUFU | gpt-3.5-turbo | 48 | 0 | 9624 | 786 | 3.920199501 | 0.00 |
| TAF1 | gpt-3.5-turbo | 336 | 0 | 9624 | 786 | 27.44139650 | 0.00 |
| TBC1D23 | gpt-3.5-turbo | 72 | 0 | 9624 | 786 | 5.880299251 | 0.00 |
| TBCK | gpt-3.5-turbo | 24 | 0 | 9624 | 786 | 1.960099750 | 0.00 |
| TBX3 | gpt-3.5-turbo | 24 | 0 | 9624 | 786 | 1.960099750 | 0.00 |
| TBXAS1 | gpt-3.5-turbo | 24 | 0 | 9624 | 786 | 1.960099750 | 0.00 |
| TCOF1 | gpt-3.5-turbo | 24 | 0 | 9624 | 786 | 1.960099750 | 0.00 |
| TMEM231 | gpt-3.5-turbo | 24 | 0 | 9624 | 786 | 1.960099750 | 0.00 |
| TMEM87B | gpt-3.5-turbo | 48 | 0 | 9624 | 786 | 3.920199501 | 0.00 |
| TUBB3 | gpt-3.5-turbo | 24 | 0 | 9624 | 786 | 1.960099750 | 0.00 |
| UNC80 | gpt-3.5-turbo | 24 | 0 | 9624 | 786 | 1.960099750 | 0.00 |
| VARS1 | gpt-3.5-turbo | 72 | 0 | 9624 | 786 | 5.880299251 | 0.00 |
| WDR26 | gpt-3.5-turbo | 624 | 0 | 9624 | 786 | 50.96259351 | 0.00 |
| WDR45 | gpt-3.5-turbo | 24 | 0 | 9624 | 786 | 1.960099750 | 0.00 |
| WFS1 | gpt-3.5-turbo | 24 | 0 | 9624 | 786 | 1.960099750 | 0.00 |
| WWOX | gpt-3.5-turbo | 24 | 0 | 9624 | 786 | 1.960099750 | 0.00 |
| YWHAG | gpt-3.5-turbo | 288 | 0 | 9624 | 786 | 23.52119700 | 0.00 |
| ZMYND11 | gpt-3.5-turbo | 48 | 0 | 9624 | 786 | 3.920199501 | 0.00 |

| Metrics | Model | Input type | Prompts | Top | Average | SD |
|---|---|---|---|---|---|---|
| Accuracy | GPT-4 | HPO | Original | Top 10 | 14.62% | 0.35 |
| Accuracy | GPT-4 | HPO | Original | Top 50 | 20.43% | 0.4 |
| Accuracy | GPT-4 | HPO | Original + Role | Top 10 | 15.80% | 0.36 |
| Accuracy | GPT-4 | HPO | Original + Role | Top 50 | 23.88% | 0.43 |
| Accuracy | GPT-4 | HPO | Original + Instruction | Top 10 | 15.34% | 0.36 |
| Accuracy | GPT-4 | HPO | Original + Instruction | Top 50 | 18.26% | 0.39 |
| Accuracy | GPT-4 | HPO | Original + Role + Instruction | Top 10 | 14.49% | 0.35 |
| Accuracy | GPT-4 | HPO | Original + Role + Instruction | Top 50 | 20.67% | 0.41 |
| Accuracy | GPT-4 | Free-text | Original | Top 10 | 12.10% | 0.33 |
| Accuracy | GPT-4 | Free-text | Original | Top 50 | 12.61% | 0.33 |
| Accuracy | GPT-4 | Free-text | Original + Role | Top 10 | 11.68% | 0.32 |
| Accuracy | GPT-4 | Free-text | Original + Role | Top 50 | 12.67% | 0.33 |
| Accuracy | GPT-4 | Free-text | Original + Instruction | Top 10 | 11.56% | 0.32 |
| Accuracy | GPT-4 | Free-text | Original + Instruction | Top 50 | 12.15% | 0.33 |
| Accuracy | GPT-4 | Free-text | Original + Role + Instruction | Top 10 | 12.67% | 0.33 |
| Accuracy | GPT-4 | Free-text | Original + Role + Instruction | Top 50 | 12.57% | 0.33 |
| Accuracy | GPT-3.5 | HPO | Original | Top 10 | 6.67% | 0.25 |
| Accuracy | GPT-3.5 | HPO | Original | Top 50 | 17.92% | 0.39 |
| Accuracy | GPT-3.5 | HPO | Original + Role | Top 10 | 6.47% | 0.25 |
| Accuracy | GPT-3.5 | HPO | Original + Role | Top 50 | 16.64% | 0.37 |
| Accuracy | GPT-3.5 | HPO | Original + Instruction | Top 10 | 11.15% | 0.31 |
| Accuracy | GPT-3.5 | HPO | Original + Instruction | Top 50 | 17.02% | 0.38 |
| Accuracy | GPT-3.5 | HPO | Original + Role + Instruction | Top 10 | 10.81% | 0.31 |
| Accuracy | GPT-3.5 | HPO | Original + Role + Instruction | Top 50 | 19.51% | 0.4 |
| Accuracy | GPT-3.5 | Free-text | Original | Top 10 | 7.69% | 0.27 |
| Accuracy | GPT-3.5 | Free-text | Original | Top 50 | 13.16% | 0.34 |
| Accuracy | GPT-3.5 | Free-text | Original + Role | Top 10 | 8.61% | 0.28 |
| Accuracy | GPT-3.5 | Free-text | Original + Role | Top 50 | 14.61% | 0.35 |
| Accuracy | GPT-3.5 | Free-text | Original + Instruction | Top 10 | 9.17% | 0.29 |
| Accuracy | GPT-3.5 | Free-text | Original + Instruction | Top 50 | 14.03% | 0.35 |
| Accuracy | GPT-3.5 | Free-text | Original + Role + Instruction | Top 10 | 8.87% | 0.28 |
| Accuracy | GPT-3.5 | Free-text | Original + Role + Instruction | Top 50 | 12.23% | 0.33 |
| Overall accuracy | GPT-4 | HPO | Original | Top 10 | 14.37% | 0.35 |
| Overall accuracy | GPT-4 | HPO | Original | Top 50 | 17.27% | 0.38 |
| Overall accuracy | GPT-4 | HPO | Original + Role | Top 10 | 15.34% | 0.36 |
| Overall accuracy | GPT-4 | HPO | Original + Role | Top 50 | 18.60% | 0.39 |
| Overall accuracy | GPT-4 | HPO | Original + Instruction | Top 10 | 15.34% | 0.36 |
| Overall accuracy | GPT-4 | HPO | Original + Instruction | Top 50 | 17.51% | 0.38 |
| Overall accuracy | GPT-4 | HPO | Original + Role + Instruction | Top 10 | 14.49% | 0.35 |
| Overall accuracy | GPT-4 | HPO | Original + Role + Instruction | Top 50 | 19.32% | 0.39 |
| Overall accuracy | GPT-4 | Free-text | Original | Top 10 | 12.00% | 0.32 |
| Overall accuracy | GPT-4 | Free-text | Original | Top 50 | 11.47% | 0.32 |
| Overall accuracy | GPT-4 | Free-text | Original + Role | Top 10 | 11.47% | 0.32 |
| Overall accuracy | GPT-4 | Free-text | Original + Role | Top 50 | 11.47% | 0.32 |
| Overall accuracy | GPT-4 | Free-text | Original + Instruction | Top 10 | 11.47% | 0.32 |
| Overall accuracy | GPT-4 | Free-text | Original + Instruction | Top 50 | 11.73% | 0.32 |
| Overall accuracy | GPT-4 | Free-text | Original + Role + Instruction | Top 10 | 12.53% | 0.33 |
| Overall accuracy | GPT-4 | Free-text | Original + Role + Instruction | Top 50 | 12.00% | 0.32 |
| Overall accuracy | GPT-3.5 | HPO | Original | Top 10 | 1.45% | 0.12 |
| Overall accuracy | GPT-3.5 | HPO | Original | Top 50 | 2.29% | 0.15 |
| Overall accuracy | GPT-3.5 | HPO | Original + Role | Top 10 | 3.99% | 0.2 |
| Overall accuracy | GPT-3.5 | HPO | Original + Role | Top 50 | 11.96% | 0.32 |
| Overall accuracy | GPT-3.5 | HPO | Original + Instruction | Top 10 | 8.57% | 0.28 |
| Overall accuracy | GPT-3.5 | HPO | Original + Instruction | Top 50 | 11.71% | 0.32 |
| Overall accuracy | GPT-3.5 | HPO | Original + Role + Instruction | Top 10 | 10.75% | 0.31 |
| Overall accuracy | GPT-3.5 | HPO | Original + Role + Instruction | Top 50 | 17.39% | 0.38 |
| Overall accuracy | GPT-3.5 | Free-text | Original | Top 10 | 4.00% | 0.2 |
| Overall accuracy | GPT-3.5 | Free-text | Original | Top 50 | 2.67% | 0.16 |
| Overall accuracy | GPT-3.5 | Free-text | Original + Role | Top 10 | 5.60% | 0.23 |
| Overall accuracy | GPT-3.5 | Free-text | Original + Role | Top 50 | 8.53% | 0.28 |
| Overall accuracy | GPT-3.5 | Free-text | Original + Instruction | Top 10 | 8.53% | 0.28 |
| Overall accuracy | GPT-3.5 | Free-text | Original + Instruction | Top 50 | 10.40% | 0.31 |
| Overall accuracy | GPT-3.5 | Free-text | Original + Role + Instruction | Top 10 | 8.80% | 0.28 |
| Overall accuracy | GPT-3.5 | Free-text | Original + Role + Instruction | Top 50 | 10.67% | 0.31 |
| Completion rate | GPT-4 | HPO | Original | Top 10 | 98.31% | 0.13 |
| Completion rate | GPT-4 | HPO | Original | Top 50 | 84.54% | 0.36 |
| Completion rate | GPT-4 | HPO | Original + Role | Top 10 | 97.10% | 0.17 |
| Completion rate | GPT-4 | HPO | Original + Role | Top 50 | 77.90% | 0.42 |
| Completion rate | GPT-4 | HPO | Original + Instruction | Top 10 | 100% | 0 |
| Completion rate | GPT-4 | HPO | Original + Instruction | Top 50 | 95.89% | 0.2 |
| Completion rate | GPT-4 | HPO | Original + Role + Instruction | Top 10 | 100% | 0 |
| Completion rate | GPT-4 | HPO | Original + Role + Instruction | Top 50 | 93.48% | 0.25 |
| Completion rate | GPT-4 | Free-text | Original | Top 10 | 99.20% | 0.09 |
| Completion rate | GPT-4 | Free-text | Original | Top 50 | 90.93% | 0.29 |
| Completion rate | GPT-4 | Free-text | Original + Role | Top 10 | 98.13% | 0.14 |
| Completion rate | GPT-4 | Free-text | Original + Role | Top 50 | 89.87% | 0.3 |
| Completion rate | GPT-4 | Free-text | Original + Instruction | Top 10 | 99.20% | 0.09 |
| Completion rate | GPT-4 | Free-text | Original + Instruction | Top 50 | 96.53% | 0.18 |
| Completion rate | GPT-4 | Free-text | Original + Role + Instruction | Top 10 | 98.93% | 0.1 |
| Completion rate | GPT-4 | Free-text | Original + Role + Instruction | Top 50 | 95.47% | 0.21 |
| Completion rate | GPT-3.5 | HPO | Original | Top 10 | 21.74% | 0.41 |
| Completion rate | GPT-3.5 | HPO | Original | Top 50 | 12.80% | 0.33 |
| Completion rate | GPT-3.5 | HPO | Original + Role | Top 10 | 61.59% | 0.49 |
| Completion rate | GPT-3.5 | HPO | Original + Role | Top 50 | 71.86% | 0.45 |
| Completion rate | GPT-3.5 | HPO | Original + Instruction | Top 10 | 76.93% | 0.42 |
| Completion rate | GPT-3.5 | HPO | Original + Instruction | Top 50 | 68.84% | 0.46 |
| Completion rate | GPT-3.5 | HPO | Original + Role + Instruction | Top 10 | 99.40% | 0.08 |
| Completion rate | GPT-3.5 | HPO | Original + Role + Instruction | Top 50 | 89.13% | 0.31 |
| Completion rate | GPT-3.5 | Free-text | Original | Top 10 | 52.00% | 0.5 |
| Completion rate | GPT-3.5 | Free-text | Original | Top 50 | 20.27% | 0.4 |
| Completion rate | GPT-3.5 | Free-text | Original + Role | Top 10 | 65.07% | 0.48 |
| Completion rate | GPT-3.5 | Free-text | Original + Role | Top 50 | 58.40% | 0.49 |
| Completion rate | GPT-3.5 | Free-text | Original + Instruction | Top 10 | 93.07% | 0.25 |
| Completion rate | GPT-3.5 | Free-text | Original + Instruction | Top 50 | 74.13% | 0.44 |
| Completion rate | GPT-3.5 | Free-text | Original + Role + Instruction | Top 10 | 99.20% | 0.09 |
| Completion rate | GPT-3.5 | Free-text | Original + Role + Instruction | Top 50 | 87.20% | 0.33 |
| Compliance rate | GPT-4 | HPO | Original | Top 10 | 100% | 0 |
| Compliance rate | GPT-4 | HPO | Original | Top 50 | 56.88% | 0.5 |
| Compliance rate | GPT-4 | HPO | Original + Role | Top 10 | 100% | 0 |
| Compliance rate | GPT-4 | HPO | Original + Role | Top 50 | 62.32% | 0.48 |
| Compliance rate | GPT-4 | HPO | Original + Instruction | Top 10 | 100% | 0 |
| Compliance rate | GPT-4 | HPO | Original + Instruction | Top 50 | 61.35% | 0.49 |
| Compliance rate | GPT-4 | HPO | Original + Role + Instruction | Top 10 | 100% | 0 |
| Compliance rate | GPT-4 | HPO | Original + Role + Instruction | Top 50 | 59.54% | 0.49 |
| Compliance rate | GPT-4 | Free-text | Original | Top 10 | 100% | 0 |
| Compliance rate | GPT-4 | Free-text | Original | Top 50 | 51.73% | 0.5 |
| Compliance rate | GPT-4 | Free-text | Original + Role | Top 10 | 100% | 0 |
| Compliance rate | GPT-4 | Free-text | Original + Role | Top 50 | 59.47% | 0.49 |
| Compliance rate | GPT-4 | Free-text | Original + Instruction | Top 10 | 100% | 0 |
| Compliance rate | GPT-4 | Free-text | Original + Instruction | Top 50 | 55.47% | 0.5 |
| Compliance rate | GPT-4 | Free-text | Original + Role + Instruction | Top 10 | 99.73% | 0.05 |
| Compliance rate | GPT-4 | Free-text | Original + Role + Instruction | Top 50 | 54.93% | 0.5 |
| Compliance rate | GPT-3.5 | HPO | Original | Top 10 | 78.14% | 0.41 |
| Compliance rate | GPT-3.5 | HPO | Original | Top 50 | 64.49% | 0.48 |
| Compliance rate | GPT-3.5 | HPO | Original + Role | Top 10 | 39.13% | 0.49 |
| Compliance rate | GPT-3.5 | HPO | Original + Role | Top 50 | 20.29% | 0.4 |
| Compliance rate | GPT-3.5 | HPO | Original + Instruction | Top 10 | 22.34% | 0.42 |
| Compliance rate | GPT-3.5 | HPO | Original + Instruction | Top 50 | 16.55% | 0.37 |
| Compliance rate | GPT-3.5 | HPO | Original + Role + Instruction | Top 10 | 0.24% | 0.05 |
| Compliance rate | GPT-3.5 | HPO | Original + Role + Instruction | Top 50 | 0% | 0 |
| Compliance rate | GPT-3.5 | Free-text | Original | Top 10 | 48.27% | 0.5 |
| Compliance rate | GPT-3.5 | Free-text | Original | Top 50 | 40.80% | 0.49 |
| Compliance rate | GPT-3.5 | Free-text | Original + Role | Top 10 | 35.20% | 0.48 |
| Compliance rate | GPT-3.5 | Free-text | Original + Role | Top 50 | 29.87% | 0.46 |
| Compliance rate | GPT-3.5 | Free-text | Original + Instruction | Top 10 | 6.93% | 0.25 |
| Compliance rate | GPT-3.5 | Free-text | Original + Instruction | Top 50 | 6.40% | 0.25 |
| Compliance rate | GPT-3.5 | Free-text | Original + Role + Instruction | Top 10 | 0.80% | 0.09 |
| Compliance rate | GPT-3.5 | Free-text | Original + Role + Instruction | Top 50 | 0.27% | 0.05 |
| Accuracy | llama2-7b-chat | HPO | Original + Role + Instruction | Top 10 | 5.41% | 0.23 |
| Accuracy | llama2-7b-chat | HPO | Original + Role + Instruction | Top 50 | 5.39% | 0.23 |
| Accuracy | llama2-7b-chat | Free-text | Original + Role + Instruction | Top 10 | 5.20% | 0.22 |
| Accuracy | llama2-7b-chat | Free-text | Original + Role + Instruction | Top 50 | 3.17% | 0.18 |
| Accuracy | llama2-13b-chat | HPO | Original + Role + Instruction | Top 10 | 7.41% | 0.26 |
| Accuracy | llama2-13b-chat | HPO | Original + Role + Instruction | Top 50 | 10.88% | 0.31 |
| Accuracy | llama2-13b-chat | Free-text | Original + Role + Instruction | Top 10 | 6.61% | 0.25 |
| Accuracy | llama2-13b-chat | Free-text | Original + Role + Instruction | Top 50 | 4.19% | 0.2 |
| Accuracy | llama2-70b-chat | HPO | Original + Role + Instruction | Top 10 | 8.21% | 0.27 |
| Accuracy | llama2-70b-chat | HPO | Original + Role + Instruction | Top 50 | 9.09% | 0.29 |
| Accuracy | llama2-70b-chat | Free-text | Original + Role + Instruction | Top 10 | 5.49% | 0.23 |
| Accuracy | llama2-70b-chat | Free-text | Original + Role + Instruction | Top 50 | 5.52% | 0.23 |
| Overall accuracy | llama2-7b-chat | HPO | Original + Role + Instruction | Top 10 | 4.71% | 0.21 |
| Overall accuracy | llama2-7b-chat | HPO | Original + Role + Instruction | Top 50 | 3.02% | 0.17 |
| Overall accuracy | llama2-7b-chat | Free-text | Original + Role + Instruction | Top 10 | 4.53% | 0.21 |
| Overall accuracy | llama2-7b-chat | Free-text | Original + Role + Instruction | Top 50 | 1.60% | 0.13 |
| Overall accuracy | llama2-13b-chat | HPO | Original + Role + Instruction | Top 10 | 6.64% | 0.25 |
| Overall accuracy | llama2-13b-chat | HPO | Original + Role + Instruction | Top 50 | 5.68% | 0.23 |
| Overall accuracy | llama2-13b-chat | Free-text | Original + Role + Instruction | Top 10 | 6.40% | 0.24 |
| Overall accuracy | llama2-13b-chat | Free-text | Original + Role + Instruction | Top 50 | 2.40% | 0.15 |
| Overall accuracy | llama2-70b-chat | HPO | Original + Role + Instruction | Top 10 | 7.13% | 0.26 |
| Overall accuracy | llama2-70b-chat | HPO | Original + Role + Instruction | Top 50 | 7.61% | 0.27 |
| Overall accuracy | llama2-70b-chat | Free-text | Original + Role + Instruction | Top 10 | 4.00% | 0.2 |
| Overall accuracy | llama2-70b-chat | Free-text | Original + Role + Instruction | Top 50 | 4.53% | 0.21 |
| Completion rate | llama2-7b-chat | HPO | Original + Role + Instruction | Top 10 | 87.08% | 0.34 |
| Completion rate | llama2-7b-chat | HPO | Original + Role + Instruction | Top 50 | 56.04% | 0.5 |
| Completion rate | llama2-7b-chat | Free-text | Original + Role + Instruction | Top 10 | 87.20% | 0.33 |
| Completion rate | llama2-7b-chat | Free-text | Original + Role + Instruction | Top 50 | 50.40% | 0.5 |
| Completion rate | llama2-13b-chat | HPO | Original + Role + Instruction | Top 10 | 89.61% | 0.31 |
| Completion rate | llama2-13b-chat | HPO | Original + Role + Instruction | Top 50 | 52.17% | 0.5 |
| Completion rate | llama2-13b-chat | Free-text | Original + Role + Instruction | Top 10 | 96.80% | 0.18 |
| Completion rate | llama2-13b-chat | Free-text | Original + Role + Instruction | Top 50 | 57.33% | 0.5 |
| Completion rate | llama2-70b-chat | HPO | Original + Role + Instruction | Top 10 | 86.64% | 0.34 |
| Completion rate | llama2-70b-chat | HPO | Original + Role + Instruction | Top 50 | 83.70% | 0.37 |
| Completion rate | llama2-70b-chat | Free-text | Original + Role + Instruction | Top 10 | 72.80% | 0.45 |
| Completion rate | llama2-70b-chat | Free-text | Original + Role + Instruction | Top 50 | 82.13% | 0.38 |
| Compliance rate | llama2-7b-chat | HPO | Original + Role + Instruction | Top 10 | 0.12% | 0.03 |
| Compliance rate | llama2-7b-chat | HPO | Original + Role + Instruction | Top 50 | 0.85% | 0.09 |
| Compliance rate | llama2-7b-chat | Free-text | Original + Role + Instruction | Top 10 | 0% | 0 |
| Compliance rate | llama2-7b-chat | Free-text | Original + Role + Instruction | Top 50 | 0.80% | 0.09 |
| Compliance rate | llama2-13b-chat | HPO | Original + Role + Instruction | Top 10 | 0.36% | 0.06 |
| Compliance rate | llama2-13b-chat | HPO | Original + Role + Instruction | Top 50 | 1.69% | 0.13 |
| Compliance rate | llama2-13b-chat | Free-text | Original + Role + Instruction | Top 10 | 0% | 0 |
| Compliance rate | llama2-13b-chat | Free-text | Original + Role + Instruction | Top 50 | 1.87% | 0.14 |
| Compliance rate | llama2-70b-chat | HPO | Original + Role + Instruction | Top 10 | 0.12% | 0.03 |
| Compliance rate | llama2-70b-chat | HPO | Original + Role + Instruction | Top 50 | 0.24% | 0.05 |
| Compliance rate | llama2-70b-chat | Free-text | Original + Role + Instruction | Top 10 | 0.27% | 0.05 |
| Compliance rate | llama2-70b-chat | Free-text | Original + Role + Instruction | Top 50 | 0.53% | 0.07 |